\documentclass[twocolumn,trackchanges]{aastex62}

\graphicspath{{./}{figures/}}

\shorttitle{Stellar Proper Motions in the Orion Nebula Cluster}
\shortauthors{Kim et al.}

\begin{document}

\title{Stellar Proper Motions in the Orion Nebula Cluster}

\correspondingauthor{Dongwon Kim}
\email{dongwon.kim@berkeley.edu}

\author[0000-0002-6658-5908]{Dongwon Kim}
\affil{Department of Astronomy, University of California, Berkeley, Berkeley, CA, 94720-3411, USA}

\author[0000-0001-9611-0009]{Jessica R. Lu}
\affil{Department of Astronomy, University of California, Berkeley, Berkeley, CA, 94720-3411, USA}

\author[0000-0002-9936-6285]{Quinn Konopacky}
\affil{Department of Physics, University of California, San Diego, 9500 Gilman Dr., La Jolla, CA 92093, USA}

\author[0000-0002-1437-4463]{Laurie Chu}
\affil{Institute for Astronomy, University of Hawaii, Manoa, 640 N. Auhoku Pl. \#209, Hilo, HI 96720 USA}

\author{Elizabeth Toller}
\noaffiliation

\author[0000-0003-2861-3995]{Jay Anderson}
\affil{Space Telescope Science Institute, 3700 San Martin Drive, Baltimore MD 21218, USA}

\author[0000-0002-9807-5435]{Christopher A. Theissen}
\affil{Department of Physics, University of California, San Diego, 9500 Gilman Dr., La Jolla, CA 92093, USA}

\author[0000-0002-6753-2066]{Mark R. Morris}
\affil{Department of Physics and Astronomy, University of California, Los Angeles, Los Angeles, CA 90095-1547, USA}

\begin{abstract}

The Orion Nebula Cluster (ONC) is the nearest site of ongoing massive star formation, which allows us to study the kinematics and dynamics of the region in detail and constrain star formation theories. Using \textit{HST} ACS/WFPC2/WFC3IR and Keck II NIRC2 data, we have measured the proper motions of 701 stars within an $\sim6'\times6'$ field of view around the center of the ONC. We have found more than 10 escaping star candidates, concentrated predominantly at the core of the cluster. 
The proper motions of the bound stars are consistent with a normal distribution, albeit elongated north-south along the Orion filament, with proper-motion dispersions of $(\sigma_{\mu,\alpha^*}, \sigma_{\mu,\delta}) = (0.83\pm0.02,\,1.12\pm0.03)$ mas yr$^{-1}$ or intrinsic velocity dispersions of $(\sigma_{v,\alpha^*}, \sigma_{v,\delta}) = (1.57\pm0.04,\,2.12\pm0.06)$ km s$^{-1}$ assuming a distance of 400\,pc to the ONC. The cluster shows no evidence for tangential-to-radial anisotropy. Our velocity dispersion profile agrees with the prediction from the observed stellar + gas density profile from~\cite{DaRio2014}, indicating that the ONC is in virial equilibrium. This finding suggests that the cluster was formed with a low star formation efficiency per dynamical timescale based on comparisons with current star formation theories. 
Our survey also recovered high-velocity IR sources BN, \textit{I}, and \textit{n} in the BN/KL region. The estimated location of the first two sources $\sim500$ years ago agrees with that of the radio source \textit{I}, consistent with their proposed common origin from a multi-stellar disintegration. 
However, source \textit{n} appears to have a small proper motion and is unlikely to have been involved in the event.
\end{abstract}



\section{Introduction} \label{sec:intro}

The Orion Nebula Cluster (ONC) region provides an exquisite opportunity to probe the process of massive star and cluster formation in detail. The ONC is very massive, with stellar masses ranging between $0.1$ and $50\mathrm{M}_{\odot}$~\citep{Hillenbrand1997}. The mean age of the ONC is 2.2 Myr with a spread of a few Myr ~\citep{Reggiani2011}, which is consistent with the star formation activity lasting between 1.5 and 3.5 Myr. The ONC's close proximity ($\sim400$ pc) and high galactic latitude ($b\sim19^{\circ}$, or $\sim135$ pc from the Galactic plane) allows us to study individual protostars and the entire cluster in detail. This combination is beneficial because the foreground has low extinction~\citep[$A_{V}$ = 1.5 mag;][]{OY2000} and contains very few stars. Also, the Orion Molecular Cloud has a very large extinction up to $A_{V}$ = 50-100 mag~\citep{HC2000,Scandariato2011}, which reduces background confusion. Therefore, the stars observed in this region of the sky are mostly ONC members~\citep{JW1988}. The ONC allows us to probe the mechanisms that drive massive star and cluster formation, which remains a challenging problem in astrophysics. 

Currently, two main theories attempt to explain massive stellar birth and they mainly differ in how and when the mass is gathered to form the star.
The first model, called the Turbulent Fragmentation model, suggests that nearly the entire mass of individual protostars is gathered at a pre-stellar stage and that further fragmentation is halted due to external pressures from turbulence, radiation, and other forms of feedback ~\citep{MT2002,MT2003}. The Competitive Accretion model, alternatively poses that mass is gathered during the star formation process itself, with all protostars starting with a low mass and accreting a significant amount of their final mass as they move through the molecular cloud~\citep{Bonnell2001a,Bonnell2001b}. One way to discern which model is more applicable is to study the dynamics of star-forming regions. While the Turbulent Fragmentation model requires the turbulence to remain virial and the star formation rates per dynamical timescale to be low, the competitive accretion model favors a rapid collapse of the gas clump and highly efficient star formation~\citep{Krumholz2011,Krumholz2012}. Comparing the dynamical age of a star-forming cluster, such as the ONC, to the age spread of its stellar population may thus facilitate estimation of the star formation rates and distinguish the two models.

The dynamical properties of the stars can also have a significant impact on the star formation efficiency. Certain interactions could produce explosive outflows that provide feedback to the surrounding molecular cloud. The nature and frequency of these interactions inform our understanding of the role that feedback plays in halting star and cluster formation, expelling gas, and setting the overall star formation efficiency within a molecular cloud. Such an explosive event has been discovered in the ONC to the northwest of the well known Trapezium cluster~\citep[e.g.][]{Zapata2005,Zapata2006,Henney2007}. This region hosts the Kleinmann-Low (KL) Nebula and contains a well studied radio and infrared source known as the Becklin-Neugebauer object \citep[BN;][]{BN1967}; thus the region is referred to as the BN/KL region. Based on analysis of the gas motions, the explosion is highly energetic ($2-6 \times 10^{47}$ erg) and expelled over a very wide angle~\citep{KS1976,Gomez2008,Bally2011}, traced by molecules with a broad range of velocities~\citep[$>$100 km s$^{-1}$;][]{KS1976,FS2009,Bally2017}. The outflow is the brightest known source of shocked H$_2$ emission, with over 100 molecular bow shocks~\citep[e.g.][]{AB1993,Stolovy1998,Colgan2007}. Millimeter and submillimeter observations suggest that the event was likely driven by close dynamical interactions in a group of massive protostars, including BN and source \textit{I}, that resulted in a violent ejection of material and the formation of a compact binary or a stellar merger~\citep{BZ2005,Bally2017}.

There have been several previous studies of dynamical interactions and proper motions (PMs) within the Orion Nebula, both in the optical and radio. Originally, \citet{Parenago1954} determined PMs for stars in the Orion Nebula over a field of $\sim$ 9 square degrees. Later, a 77 year baseline survey was done by ~\citet{vanAltena1988} for 73 stars in the Orion Nebula. \citet{JW1988} then carried out a survey using deep red-optical plates taken over 23 years on the Lick Shane reflector, which included over 1000 stars within 15$\arcmin$ of the ONC. In the radio, \citet{Gomez2005} measured the PMs of 35 sources in the Orion Nebula using the Very Large Array, with additional measurements presented in \citet{Gomez2008} and \citet{Dzib2017}. Most recently, \citet{Kuhn2019} estimated the velocity dispersion of the ONC using the PMs of 50 sources in \textit{Gaia} Data Release 2~\citep[DR2;][]{GDR2overview} within $\sim10\farcm0$ of the center of the cluster. The ONC has proven a challenging environment for measuring PMs, particularly in the very center. These previous studies are limited either by their lack of precision or small sample size. 

Fortunately, we now have access to a long baseline of data on the ONC from the Hubble Space Telescope and high-resolution near-IR data from the Keck II telescope focusing on the BN/KL region. Using these data, we have increased the precision of PMs, which has allowed us to further learn about the kinematics in this nearest massive star forming area.

We present the observations and data used to construct a new PM catalog for the ONC in \S\ref{sec:observations}. The analysis process for extracting PMs for each star is detailed in \S\ref{sec:Analysis}. The results are given in \S\ref{sec:Results} followed by a discussion of how these results compare to previous studies in \S\ref{sec:Discussion}. Also in \S\ref{sec:Discussion} we briefly discuss the interaction of sources near the BN/KL region.

\begin{figure*}[ht!]
\includegraphics[scale=0.48]{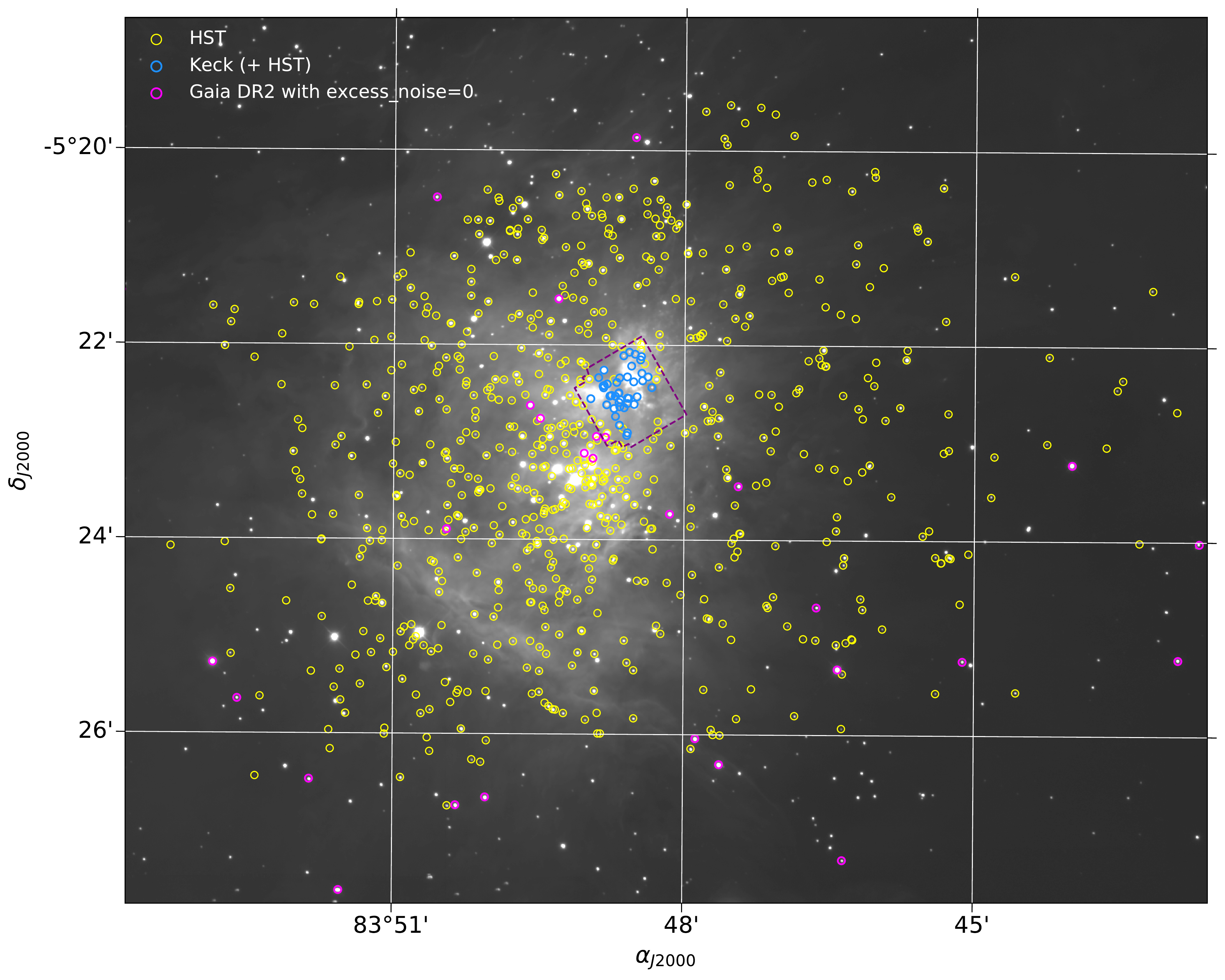}
\caption{Spatial distribution of stars in our PM catalog, overlaid on the CTIO/Blanco ISPI $K_S$-band image of the ONC from \cite{Robberto2010}. Empty yellow and blue circles mark stars measured with \textit{HST} and Keck (or Keck+\textit{HST}), respectively. Empty magenta circles mark \textit{Gaia} DR2 sources with \texttt{astrometric\_excess\_noise=0} used in \cite{Kuhn2019}. The dashed polygon illustrates the sky coverage of our 2010 and 2014 Keck NIRC2 data. 
\label{fig:mosaic}}
\end{figure*}

\section{Observations and Data} \label{sec:observations}

We measured stellar PMs near the center of the Trapezium and BN/KL region using high-resolution optical and infrared (IR) images spanning $\sim$20 years. Our final PM catalog covers $\sim6\times6$ sq arcmin around the Trapezium. The images were obtained with different instruments on board the Hubble Space Telescope (\textit{HST}) including the Advanced Camera for Surveys with the Wide Field Channel (ACS/WFC), the Wide Field Camera 3 IR detector (WFC3/IR), and the Wide Field and Planetary Camera 2 (WFPC2) as well as the Near-Infrared Camera 2 (NIRC2) of the W. M. Keck II 10-m telescope. 

\begin{deluxetable*}{ccccccccc}[ht]
\tablecaption{\textit{HST} observation \label{tab:HSTobserv}}
\tablecolumns{9}
\tablenum{1}
\tablewidth{0pt}
\tablehead{
\colhead{Epoch} & \colhead{Date} &
\colhead{$\alpha_{J2000}$} & \colhead{$\delta_{J2000}$} & \colhead{Instrument} &
\colhead{Filter} & \colhead{Exp Time} & \colhead{Proposal ID} & \colhead{PI Name} \\
\colhead{} & \colhead{(YYYY-mm-dd)} &
\colhead{(hms)} & \colhead{(dms)} & \colhead{} & 
\colhead{(nm)} & \colhead{(s)} & \colhead{} & \colhead{}
}
\startdata
1 & 1995-12-15 & 5:35:15.45 & -5:24:06.65 & WFPC2 & F547M & 200.0 & 6056 & Rubin \\
 & \dots & \dots & \dots & \dots & \dots & \dots & \dots & \dots\\ 
 & 1995-10-03 & 5:35:13.79 & -5:21:47.13 & WFPC2 & F791W & 100.0 & 5976 & O'Dell \\
 & \dots & \dots & \dots & \dots & \dots & \dots & \dots & \dots\\\hline
2  & 1998-11-02 & 5:35:00.46 & -5:24:40.00 & WFPC2 & F547M & 500.0 & 6666 & Stauffer \\
 & \dots & \dots & \dots & \dots & \dots & \dots & \dots & \dots\\
 & 1998-11-02 & 5:35:00.46 & -5:24:40.00 & WFPC2 & F791W & 300.0 & 6666 & Stauffer \\
 & \dots & \dots & \dots & \dots & \dots & \dots & \dots & \dots\\\hline
3 & 2000-09-13 & 5:35:13.77 & -5:21:47.14 & WFPC2 & F547M & 50.0 & 8121 & O'Dell \\
 & \dots & \dots & \dots & \dots & \dots & \dots & \dots & \dots\\\hline
4  & 2001-03-13 & 5:35:17.00 & -5:23:27.00 & WFPC2 & F439W & 160.0 & 8894 & Beckwith \\
 & \dots & \dots & \dots & \dots & \dots & \dots & \dots & \dots\\\hline
5  & 2004-10-12 & 5:35:18.43 & -5:22:12.62 & ACS & F435W & 420.0 & 10246 & Robberto \\
 & \dots & \dots & \dots & \dots & \dots & \dots & \dots & \dots\\
 & 2004-10-12 & 5:35:18.43 & -5:22:12.62 & ACS & F555W & 385.0 & 10246 & Robberto \\
 & \dots & \dots & \dots & \dots & \dots & \dots & \dots & \dots\\
 & 2004-10-12 & 5:35:18.43 & -5:22:12.62 & ACS & F775W & 385.0 & 10246 & Robberto \\
 & \dots & \dots & \dots & \dots & \dots & \dots & \dots & \dots\\\hline
6  & 2005-04-05 & 5:34:56.37 & -5:23:19.89 & ACS & F435W & 420.0 & 10246 & Robberto \\
 & \dots & \dots & \dots & \dots & \dots & \dots & \dots & \dots\\
 & 2005-04-05 & 5:34:56.37 & -5:23:19.89 & ACS & F555W & 385.0 & 10246 & Robberto \\
 & \dots & \dots & \dots & \dots & \dots & \dots & \dots & \dots\\
 & 2005-04-05 & 5:34:56.37 & -5:23:19.89 & ACS & F775W & 385.0 & 10246 & Robberto \\
 & \dots & \dots & \dots & \dots & \dots & \dots & \dots & \dots\\\hline
7 & 2007-09-12 & 5:35:12.05 & -5:23:27.00 & WFPC2 & F791W & 40.0 & 11038 & Biretta \\
 & \dots & \dots & \dots & \dots & \dots & \dots & \dots & \dots\\
 & 2007-09-12 & 5:35:12.05 & -5:23:27.00 & WFPC2 & F547M & 40.0 & 11038 & Biretta \\
 & \dots & \dots & \dots & \dots & \dots & \dots & \dots & \dots\\\hline
8 & 2015-02-25 & 5:34:56.37 & -5:23:19.89 & ACS & F775W & 340.0 & 13826 & Robberto \\
 & \dots & \dots & \dots & \dots & \dots & \dots & \dots & \dots\\\hline
9 & 2015-03-11 & 5:34:47.07 & -5:17:29.31 & WFC3 & F130N & 302.9 & 13826 & Robberto \\
 & \dots & \dots & \dots & \dots & \dots & \dots & \dots & \dots\\
 & 2015-03-11 & 5:34:47.07 & -5:17:29.31 & WFC3 & F139M & 302.9 & 13826 & Robberto \\
 & \dots & \dots & \dots & \dots & \dots & \dots & \dots & \dots\\\hline
10 & 2015-10-23 & 5:35:18.43 & -5:22:12.62 & ACS & F775W & 340.0 & 13826 & Robberto \\
 & \dots & \dots & \dots & \dots & \dots & \dots & \dots & \dots\\\hline
11 & 2015-10-24 & 5:35:09.34 & -5:29:55.00 & WFC3 & F130N & 302.9 & 13826 & Robberto \\
 & \dots & \dots & \dots & \dots & \dots & \dots & \dots & \dots\\
 & 2015-10-24 & 5:35:35.72 & -5:31:04.42 & WFC3 & F139M & 302.9 & 13826 & Robberto \\
 & \dots & \dots & \dots & \dots & \dots & \dots & \dots & \dots\\
\enddata
\tablecomments{Table~\ref{tab:HSTobserv} is published in its entirety in the machine-readable format. A portion is shown here for guidance regarding its form and content.}
\end{deluxetable*}

\begin{deluxetable*}{cccCrcccc}[ht!]
\tablecaption{Keck AO NIRC2 Wide-field Camera Observation \label{tab:NIRC2observ}}
\tablecolumns{10}
\tablenum{2}
\tablewidth{0pt}
\tablehead{
\colhead{Epoch} &
\colhead{Date} &
\colhead{Filter} &
\colhead{Exp Time} & \colhead{N$_{indi}$} & \colhead{N$_{stack}$} & \colhead{Total int. time} &
\colhead{FWHM} & \colhead{Strehl} \\
\colhead{} &
\colhead{(YYYY-mm-dd)} & \colhead{} &
\colhead{(s)} & \colhead{} & \colhead{} & \colhead{(s)} & \colhead{(mas)} & \colhead{}
}
\startdata
12 & 2010-10-31 & He I B & 27.15 & 53 & 9 & 1439 & 162.32 & 0.136\\ 
13 & 2014-12-11 & He I B & 27.15 & 133 & 9 & 36111 & 132.40 & 0.191\\
\enddata
\end{deluxetable*}

\subsection{HST}
The observations from \textit{HST} consisted of 11 epochs between 1995 and 2015 \citep{Rubin1997,O'Dell1997,Prosser1994,O'Dell2001,Robberto2004,Robberto2013}, mostly with medium or wide optical/IR passband filters (F435W, F439W, F539M, F555W, F775W, F791W, F139M) except for an IR filter F130N. All \textit{HST} archival images having central coordinates within $\sim3\farcm0$ of the center of the ONC were selected. However, only those images with exposure times longer than 40 seconds were used in our PM analysis to ensure sufficiently high signal-to-noise ratios for faint sources. A few of these images were also rejected in the process of matching and alignment (see Section~\ref{sec:proper motions}).

The \textit{HST} images were obtained from several different cameras. The ACS/WFC consists of two 2048$\times$4096 pixel CCD detectors. The plate scale is 50 mas per pixel, which corresponds to a $202\arcsec\times200\arcsec$ field of view. The WFPC2 uses four 800$\times$800 pixel CCDs where three of them cover a 150\arcsec$\times$150\arcsec region (WF) and have a pixel scale of 100 mas pixel$^{-1}$. The fourth CCD (PC) images a $34\arcsec\times34\arcsec$ square field with a spatial scale of 56 mas pixel$^{-1}$. The WFC3 IR channel uses a single 1024$\times$1024 pixel CCD detector with a plate scale of 130 mas pixel$^{-1}$, corresponding to a $136\arcsec\times123\arcsec$ field of view.

Observations that were within $\sim 1$ month and with the same instrument were combined to define a single epoch. In Table~\ref{tab:HSTobserv}, we provide the complete list of \textit{HST} observations for the different epochs used in this work including the epoch number, the date of the observations, the R.A.\ and decl.\ at the center of the frames, the instrument, the filter, the total exposure time, and the Principal Investigator for the data.

\subsection{Keck AO}
The observations with NIRC2 (instrument PI: K. Matthews) focused on the BN/KL star-forming region ($\alpha$ = 05:35:14.16, $\delta$ = $-$05:22:21.5). 
The data were obtained on 2010 October 30-November 1 and 2014 December 11-12 . The first run in 2010 is also described in \citet{Sitarski2013}. The observations were conducted using the laser guide star adaptive optics (LGS-AO) system~\citep{Wizinowich2006}. The laser guide star corrected for most atmospheric aberrations, however, low-order tip-tilt terms were corrected using visible light observations of the star Paranego 1839 ($\alpha_{J2000}$ = 05:35:14:64, $\delta_{J2000}$ = -05:22:33.7). In order to avoid the strong nebulosity in this region, sky frames were obtained for the wavefront sensors using larger-than-normal sky offset positions.

The two epochs of Keck AO observations covered nearly the same sky area with the wide-field cameras on NIRC2, which has a pixel scale of 39.686 mas pixel$^{-1}$~\citep{Yelda2010} and field of view of 40\arcsec$\times$40\arcsec, in the same passband, He I b ($\lambda_{0}= 2.06 \mu$m, $\Delta\lambda=0.03\mu$m). The narrow-band filter allows us to avoid the saturation of bright sources such as BN. The images were mosaicked around the BN/KL region for a total areal coverage of 1.4 square arcminutes. Sky frames were taken interspersed with science observations in a dark region $\sim15\degr$ to the east. Sky observations were timed in such a way that the field rotator mirror angle was identical to that of the science exposures, which is necessary to accurately subtract thermal emission from the field rotator mirror in this band \citep{Stolte2008}.

A summary of our Keck AO observations is listed in Table~\ref{tab:NIRC2observ}. The field of view of our Keck data is illustrated by a dashed polygon in Figure~\ref{fig:mosaic}.

\section{Analysis} \label{sec:Analysis}

\subsection{Astrometry}

\subsubsection{HST}\label{sec:HST astrometry}

For ACS/WFC and WFC3/IR data, we used pipeline-calibrated images with a suffix \texttt{\_flt}, which were dark and bias corrected and have been flat-fielded. All images were downloaded between February and June 2018 from the Mikulski Archive for Space Telescopes (MAST)\footnote{http://archive.stsci.edu}. To measure stellar positions and fluxes in each exposure, we adopted the FORTRAN code \texttt{hst1pass}\footnote{http://www.stsci.edu/$\sim$jayander/CODE/}, an advanced version of \texttt{img2xym\_WFC} software package for \textit{HST}~\citep{AK2006}. The hst1pass code runs a single pass of source finding and point-source function (PSF) fitting for each exposure and corrects the positions of stars using the geometric-distortion correction of \citet{AK2006} for ACS/WFC and the WFC3/IR correction developed by J. Anderson.\footnote{http://www.stsci.edu/$\sim$jayander/STDGDCs/} For WFPC2 data, we used calibrated images with a suffix \texttt{\_c0f} and analyzed with the FORTRAN code \texttt{img2xymrduv}~\citep{AK2003}. This code is implemented similarly to \texttt{hst1pass} and corrects the positions of stars from the WFPC2 data based on the distortion correction of \citet{AK2003}. 

Outputs from \texttt{hst1pass} and \texttt{img2xymrduv} include the distortion-corrected positions of stars, their R.A.~and decl.~based on the WCS information in the images' header, instrumental magnitudes, and the quality (or \textit{q}) of the detections. Sources with \textit{q} close to 0 appear very stellar, while those with large \textit{q} values are mostly cosmic ray impacts or artifacts of diffraction spikes. For our analysis, we apply a quality cut with the threshold of $0<q\leq0.5$ to exclude such false positives and saturated sources. We also set the minimum flux limits to be 1300 electrons for the narrow filter F130N and 500 electrons for other medium/wide filters, high enough to distinguish between the detections of stars and background noise.

\subsubsection{Keck AO}\label{sec:Keck astrometry}

The Keck AO NIRC2 data were reduced through a standard pipeline originally developed for analysis of Galactic center images \citep{Stolte2008,Lu2009}. This process includes dark and flat-field correction, sky subtraction, masking of bad pixels and cosmic rays, and the application of the distortion solution, provided by H. Fu\footnote{http://homepage.physics.uiowa.edu/$\sim$haifu/idl/nirc2wide/}. The images were then registered and drizzled using the IRAF/PyRAF modules \textit{xregister} and \textit{drizzle}. The images were all stacked into one final average image for each pointing. Additionally, each final image had three associated subimages that combined one third of the data that were used to estimate astrometric and photometric uncertainties. 

We used the IDL package \texttt{StarFinder}~\citep{Diolaiti2000} on each of the final averaged images and each of the subimages for each pointing with the wide camera to determine precise pixel positions for stars within the field. \texttt{StarFinder} extracts a PSF from the image over several iterations from a set of stars, for which we selected a total of five to seven stars in a range of magnitudes $8<K<10$.

The star catalogs from each pointing were then matched with the corresponding catalogs from their subimages with the program \texttt{align}~\citep{Ghez2008}, which crossmatches sources and solves for astrometric transformations. The final star catalogs only include stars that appeared in each of the subimages as well as the averaged images. Finally, false or low-quality detections were rejected from the final star catalogs based on the PSF correlation coefficients of individual stars provided by \texttt{StarFinder}. We applied a threshold of $corr\geq0.8$ for the quality cut, which is strict enough to reject unlikely detections while retaining faint stars~\citep{Diolaiti2000}.

\begin{deluxetable*}{ccccccccccc}[ht!]
\tablecaption{PM measurements \label{tab:PMlist}}
\tablecolumns{11}
\tablenum{3}
\tablewidth{0pt}
\tablehead{
\colhead{ID} & \colhead{$\alpha_{J2000}$\tablenotemark{a}} & \colhead{$\delta_{J2000}$\tablenotemark{a}} & \colhead{$\mu_{\alpha^*}$} & \colhead{$\epsilon_{\mu_{\alpha^*}}$} & \colhead{$\mu_\delta$} & \colhead{$\epsilon_{\mu_\delta}$} & \colhead{N$_{det}$} & \colhead{$\Delta$t} & \colhead{F139} & \colhead{Note\tablenotemark{b}} \\
\colhead{} & \colhead{deg} &
\colhead{deg} & \colhead{mas yr$^{-1}$} & \colhead{mas yr$^{-1}$} & 
\colhead{mas yr$^{-1}$} & \colhead{mas yr$^{-1}$} & \colhead{} & \colhead{yrs} & \colhead{mag} & \colhead{}
}
\startdata
1 & 83.85346518 & -5.36602620 & 0.34 & 0.09 & -0.71 & 0.05 & 4 & 16.9 & 14.57 &  \\
2 & 83.83552846 & -5.34779414 & -0.20 & 0.16 & 0.68 & 0.55 & 3 & 16.3 & 12.12 &  \\
3 & 83.77062885 & -5.35257430 & -0.43 & 0.06 & 2.12 & 0.18 & 3 & 11.0 & 13.25 &  \\
4 & 83.77579375 & -5.33816393 & -0.41 & 0.28 & -0.75 & 0.30 & 3 & 11.0 & 14.52 &  \\
5 & 83.82285598 & -5.38091090 & 1.38 & 0.16 & 1.64 & 0.05 & 5 & 16.4 & 13.35 &  \\
6 & 83.79410989 & -5.37909779 & 0.14 & 0.46 & -1.53 & 0.06 & 3 & 16.3 & 12.11 &  \\
7 & 83.81797732 & -5.34033710 & -1.00 & 0.13 & 0.01 & 0.45 & 3 & 16.3 & 12.93 &  \\
8 & 83.85596135 & -5.39258964 & -0.79 & 0.11 & -0.44 & 0.11 & 4 & 16.3 & 12.81 &  \\
9 & 83.84994021 & -5.41941865 & 0.12 & 0.81 & 0.98 & 0.15 & 4 & 14.6 & 13.04 &  \\
10 & 83.79543932 & -5.37954202 & -0.04 & 0.27 & 1.00 & 0.40 & 5 & 19.4 & 13.28 &  \\
\dots & \dots & \dots & \dots & \dots & \dots & \dots & \dots & \dots & \dots & \dots\\
\enddata
\tablenotetext{a}{Epoch 2015.5.}
\tablenotetext{b}{In this column, 1 = proplyd morphology; 2 = Herbig-Haro object~\citep{Ricci2008}; 3 = double star, previously known in the literature~\citep{Hillenbrand1997,Robberto2013}; 4 = new double star, previously reported as single in the literature; 5 = double-star candidate, previously reported as single in the literature; 6 = double-star candidate, previously unreported.}
\tablecomments{Table~\ref{tab:PMlist} is published in its entirety in the machine-readable format. A portion is shown here for guidance regarding its form and content.}
\end{deluxetable*}

\subsection{Relative PMs}\label{sec:proper motions}
 
 The first step in measuring relative PMs is to establish an astrometric reference frame. With distortion-free astrometric measurements, \textit{Gaia} DR2~\citep{Gaia2016,GDR2overview} provides an excellent foundation for a reference frame. One caveat of using \textit{Gaia} DR2 alone as a reference frame is that in the central region of ONC the photometry only reaches $G\sim17$, mostly due to high nebulosity. Given the star catalogs from different passbands, we needed a reference frame for alignment that covers a wider range of magnitudes and colors.
 
We constructed a new reference frame, building upon \textit{Gaia} DR2 using our star catalogs from ACS/WFC F775W and WFC3 IR F139M in Epochs \#8-11 as follows. First, we converted the celestial coordinates of stars in \textit{Gaia} DR 2 into right-handed Cartesian coordinates x and y parallel to the R.A. and decl. directions (i.e. x=$\Delta\alpha\cos\delta$, y=$\Delta\delta$)\footnote{Higher-order projection such as Equation (1) in~\cite{vandeVan2006} changes our PM measurements typically 1 $\mu$as yr$^{-1}$ or less.}, respectively. In order to compromise between the number of reference sources and the accuracy of the astrometry, we adopted only \textit{Gaia} stars with measurement errors smaller than 60 mas. We cross-matched the stars in the \textit{HST} F775W and F139M catalogs with those of the \textit{Gaia} DR2 catalog within a radius of 60 mas and applied 2D second-order polynomials to transform the positions of stars in the input star catalogs to those of the reference stars. Although the minimum number of data points for a second-order polynomial fit is 6, we excluded frames that had less than 9 matches to ensure a good fit. The median position was calculated for each star from both \textit{Gaia} DR2 and the transformed catalogs. During each polynomial fitting, we rejected 3-sigma outliers and repeated the same process until convergence into a final solution. 
 
In the final iteration, we used the new reference frame to transform and match all the rest of the \textit{HST} and Keck star catalogs from all epochs. A final catalog was constructed using the median of the transformed positions for each star in each epoch and the standard error of the median was adopted as the positional uncertainty. 

The relative PMs of each star were measured using least-squares straight-line fits to the positions over time. We set a lower limit for the time baseline ($\Delta$t) as 1 year. For stars identified in three epochs or more, the linear fit determined the velocities with the errors calculated using the covariance matrix from the fit. The velocities of the remaining objects, detected in only two epochs, were calculated as the positional difference between the two epochs divided by the time baseline, and their uncertainties were calculated as the quadratic sum of positional errors in each epoch. We provide our PM catalog with positions at epoch 2015.5 in Table~\ref{tab:PMlist}. To estimate the photometric depth of the catalog, we cross-matched the PM catalog with the first pass \texttt{hst1pass} outputs for F139M images, lowering the flux limit down to 20 electrons. We found that our sample includes 693 stars that fall within the range between F139M$=9.5$ and $20.5$, $\sim95$\% of which are brighter than F139M$=18.0$, and 22 stars are undetected in the near-IR passband. 

\subsection{PM dispersion calculation}\label{sec:PM dispersion calculation}

From the measured PM measurements, we derived the internal PM dispersion of the ONC using Bayes's theorem and a multivariate normal distribution model. Assuming that each PM measurement is drawn from a Gaussian distribution with mean $\bar{\mu}$ and an intrinsic dispersion $\sigma_\mu$, the likelihood for the \textit{i}th PM measurement $\mu_{i}\pm\epsilon_{i}$ is then given by
\begin{equation}
%L(\bar{\mu_{x}},\sigma_{x},\bar{\mu_{y}},\sigma_{y})=G
L_i(\mu_i\vert \bar{\mu}, \sigma_{\mu})=G(\mu_{i}\vert \bar{\mu}, \sqrt{\sigma_{\mu}^2+\epsilon_i^2}),
\end{equation}
where the final dispersion is the quadratic sum of the intrinsic dispersion, $\sigma_\mu$, and the error on the measurement for each star, $\epsilon_i$. Given a set of measurements $D\equiv\lbrace \mu_i, \epsilon_i \rbrace_{i=1}^N$, the posterior probability \textit{P} is defined by Bayes' theorem as
\begin{equation}
P(\bar{\mu},\sigma_\mu \vert D) \varpropto \prod\limits_{i} L_i(\mu_i\vert \bar{\mu}, \sigma_{\mu}) p(\bar{\mu},\sigma_{\mu}), 
\end{equation}
where $p(\bar{\mu},\sigma_\mu)\equiv p(\bar{\mu})p(\sigma_{\mu})$ is the prior for the mean and the standard deviation. We adopt a flat prior for the mean and a ``non-informative'' Jeffreys prior for the standard deviation - i.e., $p(\sigma_{\mu})\propto\sigma_{\mu}^{-1}$~\citep[see \S VII in][for justification]{Jaynes1968}\footnote{Flat prior can be strongly ``informative'' for a scale parameter and bias the posterior probability distribution~\citep[see, e.g., \S4.1 of][]{Eriksen2008}. Nevertheless, using a flat prior instead of the Jeffreys prior $\sigma_{\mu}^{-1}$ increases our estimates of PM dispersions in this paper only by $\sim0.01$ mas yr$^{-1}$ or less except for the sample in Figure~\ref{fig:difference}, by only $\sim0.05$ mas yr$^{-1}$, due to its small sample size. The choice of prior thus changes none of our concolusion in this paper.}. As each star has PM measurements along R.A. ($\alpha$) and decl.($\delta$), we maximized the product of the posteriori for the two axes, i.e. $P_{\alpha,\delta}=P_{\alpha}P_{\delta}$ utilizing the Markov chain Monte Carlo (MCMC) Ensemble sampler \texttt{emcee}~\citep{emcee}.

\begin{figure*}[ht!]
\plotone{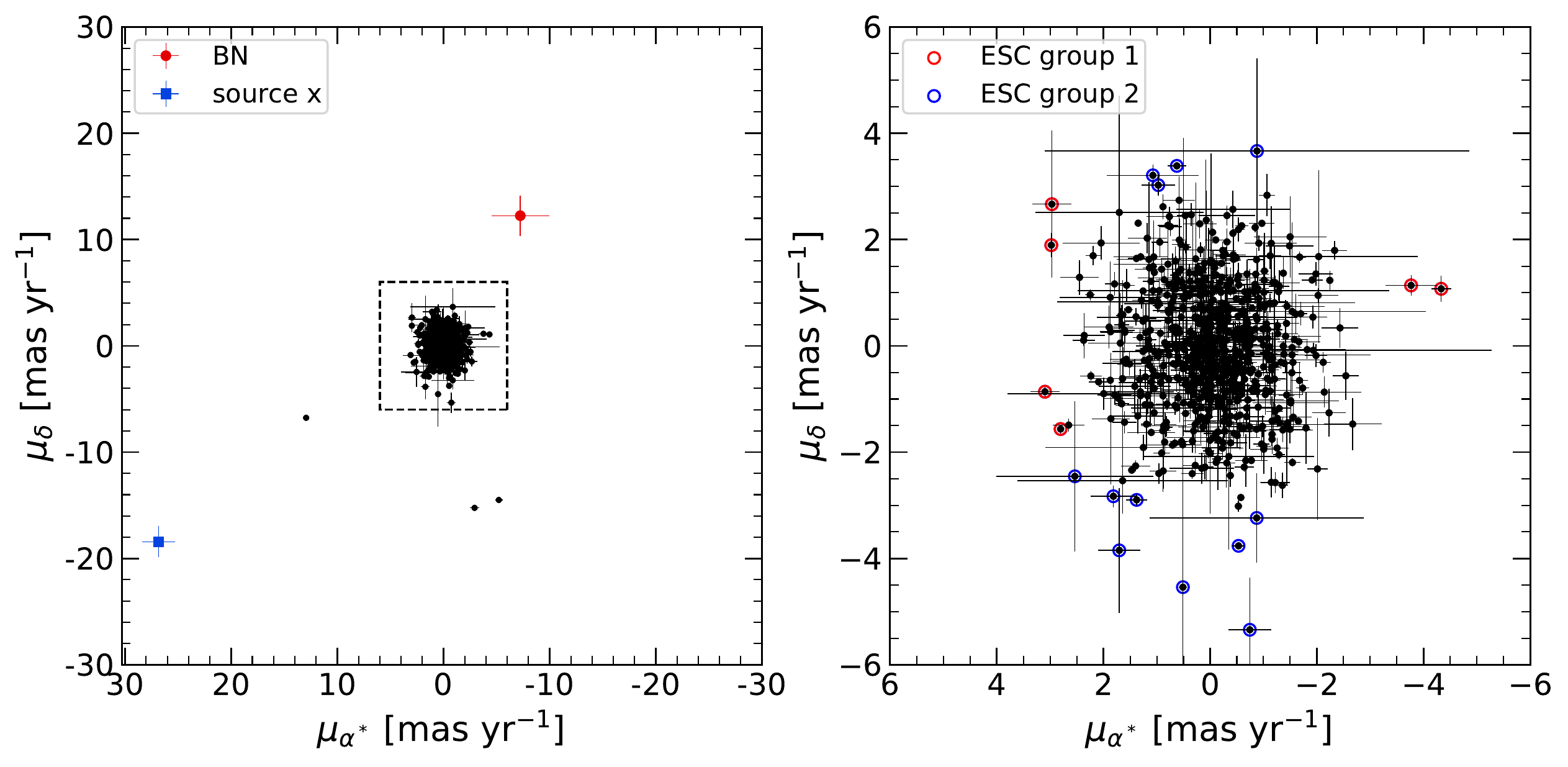}
\caption{
The PM-vector point diagram, at two different scales, for all stars measured in this work. The filled red circle and blue square in the left panel represent BN and source 'x'.  Escaping star candidates (ESCs) are marked with empty circles in the right panel (see \S \ref{sec:high_vel_stars}). 
\label{fig:VPD}}
\end{figure*}

\section{Results} \label{sec:Results}

We present the PMs for 701 stars centered on the Trapezium and the BN/KL regions (Figure~\ref{fig:VPD}), adding $\sim 500$ sources with precise PM measurements as compared to \textit{Gaia} DR2 over the same region. Our catalog has a temporal baseline of $\sim20$ years and extends the wavelength coverage to the near-IR. Throughout the remainder of the paper, we use the notations $\mu_{\alpha^*}$ and $\mu_{\delta}$ to denote projected PMs where $\mu_{\alpha^*}\equiv\mu_{\alpha}\cos\delta$.

\subsection{Consistency with Gaia DR2 PMs} \label{sec:ConsistencyWithDR2}

\begin{figure}
\plotone{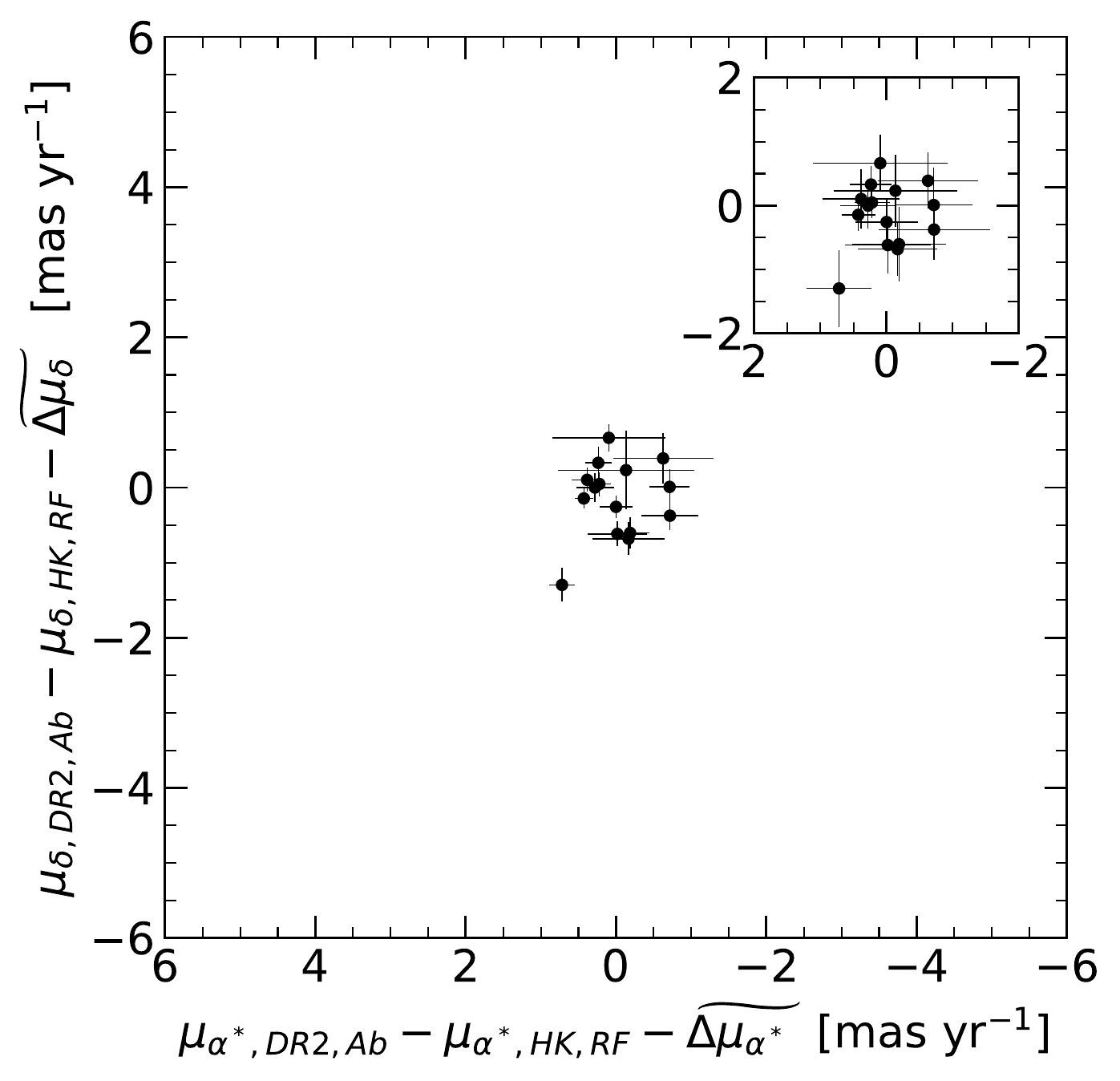}
\caption{Differences between the absolute (Ab) PMs from \textit{Gaia} DR2 and the PMs in the rest frame (RF) of the ONC based on the \textit{HST} + Keck (HK). We subtracted the median values of the differences between the absolute and the relative PMs ($\widetilde{\Delta\mu_{\alpha^*}}$, $\widetilde{\Delta\mu_{\delta}}$) = (1.6, 0.6) mas yr$^{-1}$ to take into account the bulk motion of the ONC. The inner panel is the same as the outer panel but for which the PM uncertainties of \textit{Gaia} DR2 were increased by a factor of 3 (see \S~\ref{sec:ConsistencyWithDR2}.)
\label{fig:difference}}

\end{figure}

\begin{figure}[ht!]
\plotone{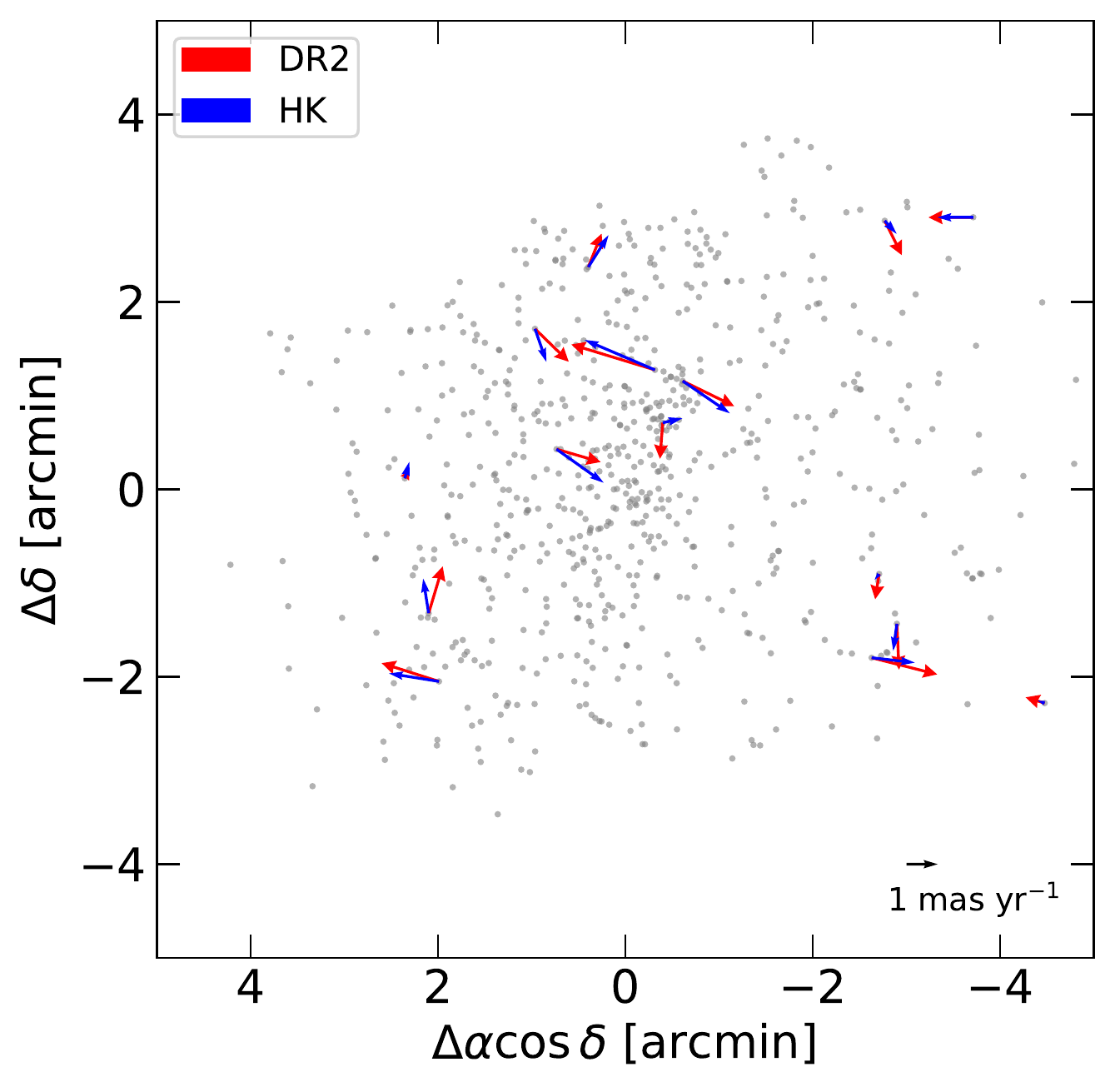}
\caption{PM vectors from \textit{Gaia} DR2 (red) and this study using \textit{HST} + Keck (HK, blue) show good agreement. In this figure, the bulk motion of the ONC inferred from the median values of the difference between the absolute and relative PMs ($\widetilde{\Delta\mu_{\alpha^*}}$, $\widetilde{\Delta\mu_{\delta}}$) = (1.6, 0.6) mas yr$^{-1}$ was subtracted from those from the \textit{Gaia} DR2 (see Figure~\ref{fig:difference}). Gray faint dots illustrate the positions of all stars in our sample. \label{fig:quiver}}
\end{figure}

We compare our PM measurements with those in \textit{Gaia} DR2. 
Due to the strong nebulosity, the astrometric measurements from \textit{Gaia} DR2 have overall lower quality for stars around the ONC compared with other nebula-free regions. As a result, we adopt a generous quality cut for \textit{Gaia} DR2 stars in order to compromise between the astrometric quality and the size of the comparison sample\footnote{The \textit{Gaia} DR2 sources with \texttt{astrometric\_excess\_noise}=0, adopted by \cite{Kuhn2019} for accurate estimates of measurement uncertainty, are generally too bright and saturated in our HST images (see Figure~\ref{fig:mosaic}).}: \texttt{astrometric\_gof\_al}$<$16 and \texttt{photometric\_mean\_g\_mag}$<$16. With this condition, we found fifteen matches between \textit{Gaia} DR2 and our PM catalog. The outer panel of Figure~\ref{fig:difference} shows the difference in PMs along the R.~A.~and decl.~axes, where the data points are concentrated around (0,0) within 1 mas yr$^{-1}$. Figure~\ref{fig:quiver} verifies that the PM vectors tend to point in similar directions with similar magnitudes. 

Returning to the outer panel of Figure~\ref{fig:difference}, however, we noticed that the differences generally exceed the measurement errors and exhibit an asymmetric distribution. The inconsistency in the amplitudes of PMs can be attributed to underestimated PM uncertainties in \textit{Gaia} DR2. \cite{Arenou2018} has demonstrated that the parallax and PM errors in \textit{Gaia} DR2 are underestimated and tend to overestimate the intrinsic dispersions for distant open/globular clusters. To test the possibility of underestimated uncertainties in \textit{Gaia} DR2 around the ONC, we compared the PM dispersions derived from the stars in common between \textit{Gaia} DR2 and our catalog, excluding kinematic outliers identified in \S\ref{sec:high_vel_stars}. 
First, we compared the PM dispersions of the fifteen stars in Figure~\ref{fig:difference}. 
The \textit{Gaia} DR2 resulted in PM dispersions ($\sigma_{\alpha^*},\sigma_{\delta}$)=($1.33^{+0.31}_{-0.23}, 0.91^{+0.21}_{-0.15}$) mas yr$^{-1}$, which is nearly 30\% larger than the dispersion from our PM measurements ($\sigma_{\alpha^*},\sigma_{\delta}$)=($1.04^{+0.26}_{-0.19}, 0.72^{+0.17}_{-0.13}$) mas yr$^{-1}$. 

In order to reach comparable PM dispersions, the PM uncertainties in \textit{Gaia} DR2 need to be increased by a factor of $\sim$3. 
Given the small size of the sample, we performed the same test using a larger sample, without the quality cut we applied for the previous sample. 
We found 197 matches between \textit{Gaia} DR2 five-parameter sources and our catalog after excluding kinematic outliers listed in \S\ref{sec:high_vel_stars} and measured the PM dispersions following the same procedure described in \S\ref{sec:dispersion}. 
We find a 2D dispersion from the \textit{Gaia} DR2 measurements of ($\sigma_{\alpha^*},\sigma_{\delta}$)=($1.50\pm0.09, 1.69\pm0.10$) mas yr$^{-1}$, which is nearly 70\% larger than the dispersion for our catalog of ($0.83\pm0.05, 1.00\pm0.06$) mas yr$^{-1}$. For comparable PM dispersions, increasing the PM uncertainties in the \textit{Gaia} by a factor of $\sim$3 is again required. We note that this increased error is still needed even when comparing to the PM dispersions for our entire sample or from previous surveys at optical wavelengths \citep[see \S\ref{sec:dispersion} and][]{JW1988}. The inner panel of Figure~\ref{fig:difference} shows the same as the outer panel but with PM uncertainties in \textit{Gaia} DR2 increased by a factor of 3, where the differences appear to be consistent overall with zero within $\sim1\sigma$. There is one outlier, whose PM difference offset toward the southeast. This star has the smallest number of both good observations and visibility periods in \textit{Gaia} DR2 out of all the matched stars, whose \texttt{astrometric\_n\_good\_obs\_al} and \texttt{visibility\_preiods\_used} values range from 106 to 205 and from 7 to 12, respectively. The outlier thus falls within the regime of possible systematic errors in PMs of \textit{Gaia} DR2 induced by the scanning law of the survey as demonstrated in Appendix A. The direction of the bias is also consistent with the scan direction around the ONC, northwest-southeast, which can be traced by the positions of \textit{Gaia} DR2 sources filtered based on the number of observations or visibility periods. We therefore conclude that, compared to \textit{Gaia} DR2, there is no significant discrepancy ascribed to our PM measurements.

\subsection{High-velocity Stars}
\label{sec:high_vel_stars}

\begin{figure*}[ht!]
\plotone{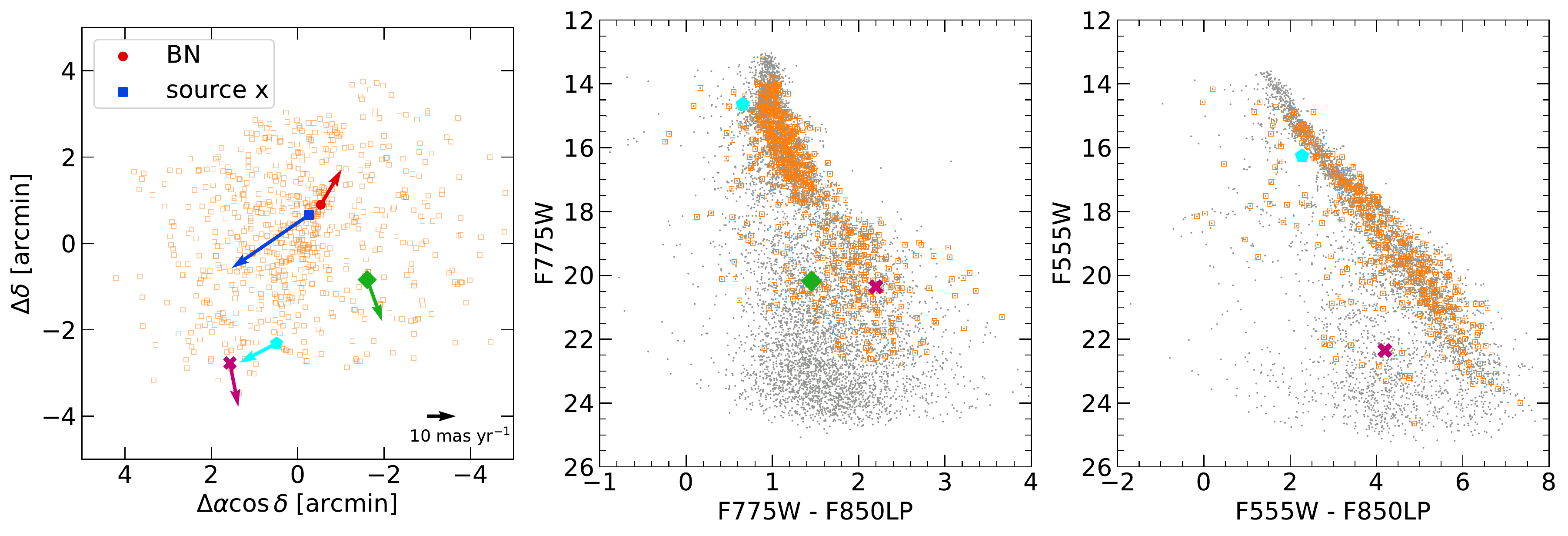}
\caption{Positions and color-magnitude distribution of stars in our sample. Left panel: the positions of all stars with PM measurements in this work (orange squares) and PM vectors for stars with large PMs. Middle panel: (F775W-F850LP) vs F775W color-magnitude diagram from \textit{HST}/ACS photometry of \cite{Robberto2013}. Previously identified members BN and source x are omitted as they are not detected at optical wavelengths. Right panel: the same as the middle panel but for (F555W-F850LP) vs F555W. The star marked as a green diamond in the left panels is omitted as it is not detected at F555W.\label{fig:CMD}}
\end{figure*}

Our analysis recovered known high-velocity stars, BN and source \textit{I}, in the BN/KL complex. We note that these sources were not detected in the optical but only in the IR images from Keck/NIRC2 and WFC3/IR. We measure PMs of ($\mu_{\alpha^*}$, $\mu_{\delta}$) = ($-7.2\pm2.7$, $12.2\pm1.9$ mas yr$^{-1}$) for BN and ($26.8\pm1.5$, $-18.4\pm1.5$ mas yr$^{-1}$) for source \textit{I}. Our measurements agree with those both at IR wavelengths from~\cite{Luhman2017} and in the radio from~\cite{Rodriguez2017} within $1\sigma$. 

We also recovered source \textit{n} in the BN/KL complex, whose PM was reported as high as $\sim7$ mas yr$^{-1}$ in some radio studies~\citep{Rodriguez2017}. In fact, the source appears highly elongated at radio wavelengths, which hinders a reliable PM measurement. At IR wavelengths, it appears as a single point source with a much smaller PM value \citep{Luhman2017}. Not surprisingly, our PM measurement for  source \textit{n} ($1.9\pm1.0$, $1.0\pm0.7$) mas yr$^{-1}$ agrees reasonably well with the motions of ($-1.8\pm1.4$, $-2.5\pm1.4$) mas yr$^{-1}$ previously measured in the IR by \citet{Luhman2017} or ($1.6\pm1.6$, $3.4\pm1.6$) mas yr$^{-1}$ in the millimeter~\citep{Goddi2011} but disagrees with the PM of ($0.0\pm0.9$, $-7.8\pm0.6$) mas yr$^{-1}$ in the radio data~\citep{Rodriguez2017}.

In addition to the previously known high-velocity stars, we detected three other stars with large PMs, as shown in the left panel of Figure~\ref{fig:VPD}. We note that we initially had a few more candidate stars with large PM values; but, after visual inspection, they were identified as false positives ascribed to marginally resolved double stars, Herbig-Haro objects, or proplyds~\citep{Prosser1994,Hillenbrand1997,Reipurth2007,Ricci2008,Robberto2013,Duchene2018}. Since foreground field stars often have large PMs, we performed further investigation to determine the nature of the three objects. Figure~\ref{fig:CMD} shows the \textit{HST}/ACS photometry of stars covering $\sim600$ arcmin$^2$ around the ONC from~\cite{Robberto2013}. The colors and magnitudes of the four objects are systematically bluer than the young ONC sequence and found in the locus of foreground objects with low reddening. This finding strengthens the idea that these kinematic outliers are most likely field stars. We note the the brightest object, a cyan pentagon in Figure~\ref{fig:CMD}, corresponds to the source \#583 in the photometric survey of ~\cite{Hillenbrand1997}, where the star was classified as a non-member with 0\% membership probability. This object also corresponds to the source \#3017360902234836608 in \textit{Gaia} DR2 with a relatively large parallax $4.14\pm0.07$ mas ($\equiv242\pm4$ pc), which supports that it is likely a foreground star. For internal kinematic analysis, we do not include these three peculiar objects. Also excluded are BN and source \textit{x} as was done in previous studies~\citep[e.g.][]{JW1988,Dzib2017}. 

\begin{figure*}[ht!]
\plotone{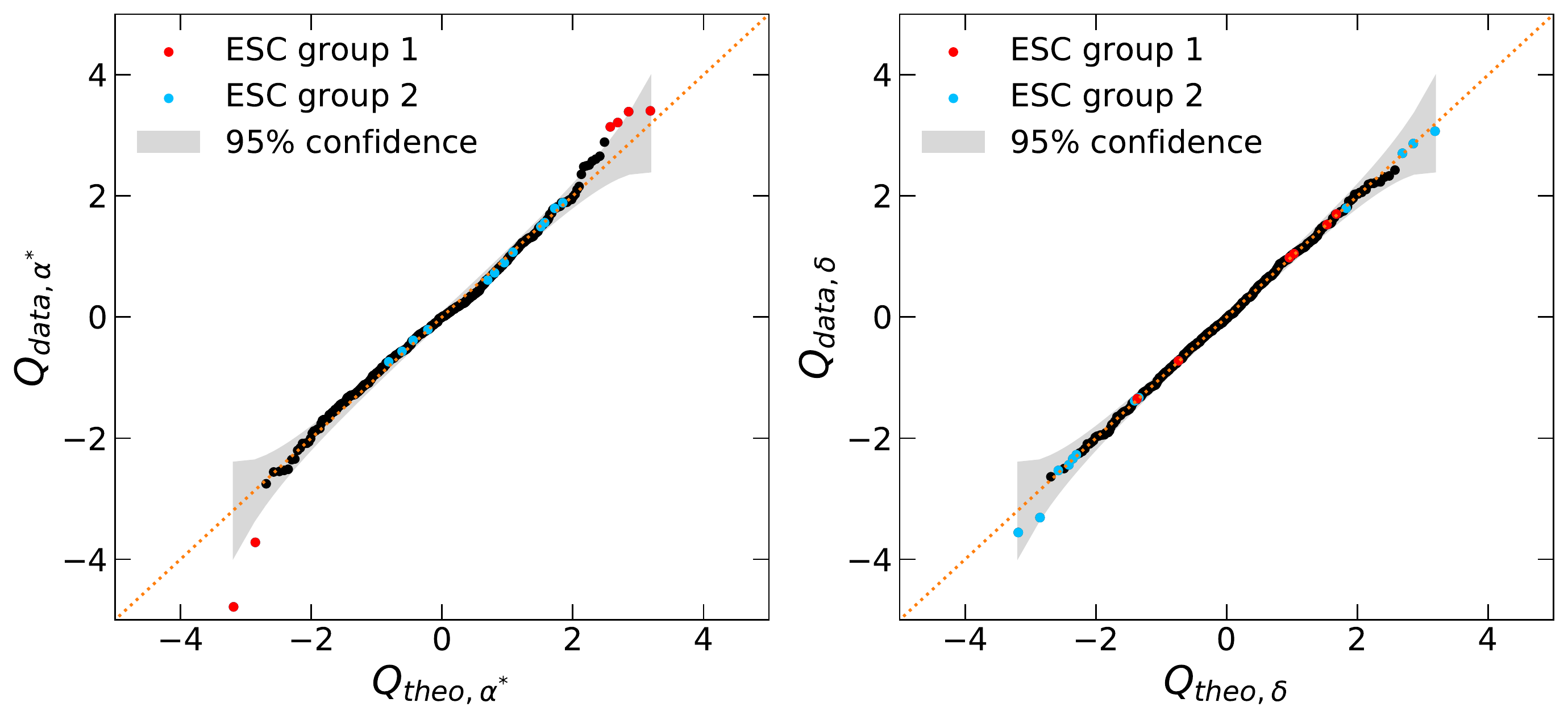}\plotone{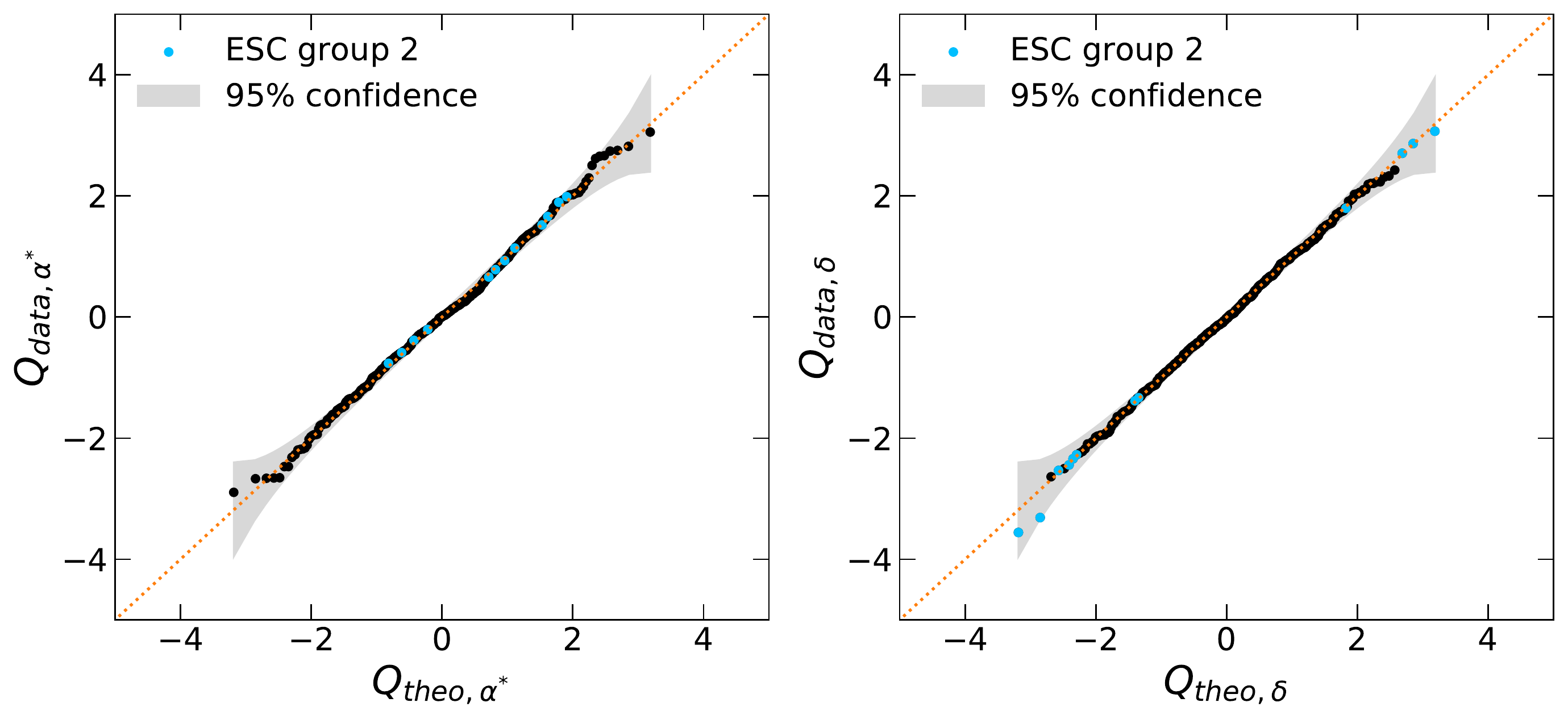}
\caption{Quantile-Quantile plots (or Q-Q plots) to assess the normality of the stellar PM distribution, comparing data quantiles to theoretical Gaussian quantiles. The dotted lines mark the expected values of the theoretical distribution. The grey regions illustrate 95\% point-wise confidence envelopes. Upper and lower panels show before and after removing the escaping star candidates, ESC group 1, from our sample. \label{fig:QQ}}
\end{figure*}

We also identify probable escaping, or evaporating, stars whose high velocities deviate significantly from our Gaussian velocity distribution model (see \S\ref{sec:isotropy}). As demonstrated by \cite{Kuhn2019}, the deviation can be visualized on a plot of observed data quantiles ($Q_{data}$) versus the theoretical quantiles of the Gaussian distribution ($Q_{theo}$). The quantiles are:
\begin{equation}
Q_{data,i}=\frac{\mu_i-\bar{\mu}}{\sqrt{\sigma^2_{\mu}+\epsilon^2_i}}
\end{equation}
\begin{equation}
Q_{theo,i}=\sqrt{2}\,\rm{erf}^{-1}(2(r_i-0.5)/n-1),
\end{equation}
where $\rm{erf^{-1}}$ is the inverse of the error function, $n$ is the number of measurements, and $r_i$ the rank of the $i$th measurement. The mean $\bar{\mu}$ and the standard deviation $\sigma_{\mu}$ are computed with the method described in \S\ref{sec:PM dispersion calculation}. The upper panels in Figure~\ref{fig:QQ} show the Q-Q plots of all stars except for the high-velocity stars. Overall, the velocity distribution is well fit by a normal distribution along both $\alpha$ and $\delta$ axes. In the $\alpha$ axis, however, we notice that beyond $Q_{data,\alpha^*}=\pm3$ the data quantiles deviate from the expected quantile function of the Gaussian distribution. In the lower panels, after excluding the 6 stars outside $Q_{data,\alpha^*}=\pm3$ (filled circles in red) and recalculating the mean and the standard deviation, the velocity distribution exhibits consistency with a normal distribution within the 95\% confidence envelope. 

We consider the possibility that the outlier stars are evaporating from the cluster by comparing their velocities to the escape speed. Using the virial theorem, the mean-square escape speed can be estimated as:
\begin{equation}
\langle v_{e}^2 \rangle^{1/2} = 2\langle v^2 \rangle^{1/2},
\end{equation}
where $\langle v^2 \rangle^{1/2}$ is the mean-square speed of the cluster's stars~\citep{BT2008}. Using this relation, we approximate a corresponding angular escape speed of $\approx3.1$ mas yr$^{-1}$. We found that the apparent angular speed of the outliers ranges from 3.2 to 4.5 mas yr$^{-1}$, all of which exceed the speed limit for evaporation. We identified 12 additional candidate stars from our sample whose apparent angular speeds exceed this threshold (see Figure~\ref{fig:VPD}), although they do not stand out as statistically significant outliers on the Q-Q plots partially due to large measurement errors or dispersion along $\delta$ axis (see Figure~\ref{fig:VPD}). We note that we initially had several more candidates that were excluded after visual inspection showed that they were marginally resolved double stars or unresolved double-star candidates with highly elongated, double-lobed morphology in \textit{HST}/ACS or Keck/NIRC2 images. 
Hereafter, we refer to the relatively high- and low-significance escaping star candidates (ESCs) as ESC group 1 and ESC group 2, respectively. In \S\ref{sec:isotropy}, we demonstrate that these stars preferentially occupy the central region of the ONC. To account for their effect on the radial variation of the velocity dispersion, we present the PM dispersion as a function of radius in Table~\ref{tab:Dispersion_distance}, for three cases: excluding (a) none of ESC groups, (b)  ESC group 1, and (c)  ESC group 1 + 2. Otherwise, we only exclude ESC group 1 when modeling the PM distribution in the following sections. 

We note that all false positives are included and flagged in our PM catalog (Table~\ref{tab:PMlist}). Amongst the false positives, we have newly identified 2 double stars and 2 double-star candidates.

\subsection{Internal PM dispersions}
\label{sec:dispersion}

Using the method described in \S\ref{sec:PM dispersion calculation}, we obtain the mean PM and intrinsic dispersion of the ONC in Cartesian coordinates:
$$\bar{\mu}_{\alpha^*}=-0.04\pm0.03 \, \rm{mas\,yr^{-1}},$$
$$\bar{\mu}_{\delta}=-0.05\pm0.05 \, \rm{mas\,yr^{-1}},$$
$$\sigma_{\mu,\alpha^*}=0.83\pm0.02 \, \rm{mas\,yr^{-1}},$$
$$\sigma_{\mu,\delta}=1.12\pm0.03 \, \rm{mas\,yr^{-1}}.$$

Following the same procedure, we also computed the mean PM and dispersion along the radial axis away from the cluster center and the tangential axis perpendicular to it:
$$\bar{\mu}_{r}=-0.04\pm0.04 \, \rm{mas\,yr^{-1}},$$
$$\bar{\mu}_{t}=0.00\pm0.04 \, \rm{mas\,yr^{-1}},$$
$$\sigma_{\mu,r}=0.97\pm0.03 \, \rm{mas\,yr^{-1}},$$
$$\sigma_{\mu,t}=1.00\pm0.03 \, \rm{mas\,yr^{-1}}.$$
Our PM dispersions agree with those found by \cite{JW1988}; ($\sigma_{\mu,\alpha^*}, \sigma_{\mu,\delta}$)=($0.91\pm0.06, 1.18\pm0.05$), ($\sigma_{\mu,r},\sigma_{\mu,t}$)=($1.06\pm0.05, 1.04\pm0.05$) mas yr$^{-1}$. 

The PM dispersions were also measured for stars grouped into equally-partitioned magnitude bins in F139M (N=110) and by distance from the center of the ONC, given in Table~\ref{tab:Dispersion_magnitude} and Table~\ref{tab:Dispersion_distance}. 
The one-dimensional PM dispersions $\sigma_{\mu,1D}$ were obtained by taking the quadratic mean of R.A. and decl. dispersions, $\sigma_{\mu,\alpha^*}$ and $\sigma_{\mu,\delta}$. The PM dispersions appear to be essentially flat within uncertainties from $m_{F139M}=$ 9.50 to 16.09, below which there is a marginal evidence of decreasing R.A. dispersion. The PM dispersions more obviously decrease with radius from the center to $R=3\farcm0$.

\begin{deluxetable*}{ccccccc}
\tablecaption{PMs as a function of magnitude \label{tab:Dispersion_magnitude}}
\tablecolumns{4}
\tablenum{4}
\tablewidth{0pt}
\tablehead{
\colhead{F139M} & \colhead{$\sigma_{\mu,\alpha^*}$} &
\colhead{$\sigma_{\mu,\delta}$} & \colhead{$\sigma_{\mu,r}$} & 
\colhead{$\sigma_{\mu,t}$} & \colhead{$\sigma_{\mu,\mathrm{1D}}$} & \colhead{N} \\
\colhead{mag} & \colhead{mas yr$^{-1}$} & \colhead{mas yr$^{-1}$} &
\colhead{mas yr$^{-1}$} & \colhead{mas yr$^{-1}$} & \colhead{mas yr$^{-1}$} & \colhead{} \\
}
\startdata
$\,\,9.50-12.18$ & $0.86\pm0.07$ & $1.08\pm0.08$ & $0.91\pm0.07$ & $1.02\pm0.08$ & $0.98\pm0.05$ & 110\\
$12.18-13.00$ & $0.87\pm0.07$ & $1.18\pm0.09$ & $1.09\pm0.08$ & $1.00\pm0.07$ & $1.04\pm0.06$ & 110\\
$13.00-13.56$ & $0.84\pm0.07$ & $1.11\pm0.09$ & $0.92\pm0.07$ & $1.05\pm0.08$ & $0.98\pm0.06$ & 110\\
$13.56-14.45$ & $0.85\pm0.07$ & $1.19\pm0.09$ & $1.01\pm0.08$ & $1.07\pm0.08$ & $1.03\pm0.06$ & 110\\
$14.45-16.09$ & $0.83\pm0.06$ & $1.12\pm0.08$ & $1.06\pm0.08$ & $0.94\pm0.07$ & $0.99\pm0.05$ & 110\\
$16.09-18.38$ & $0.67\pm0.05$ & $1.03\pm0.08$ & $0.79\pm0.06$ & $0.94\pm0.07$ & $0.87\pm0.05$ & 110\\
\enddata
\tablenote{The dispersion columns give 1D intrinsic dispersions in the R.A., decl., radial, and tangential directions. The final dispersion column is the mean of the R.A.\ and decl.\ dispersions.}
\end{deluxetable*}

\begin{deluxetable*}{cccccccc}
\tablecaption{PM dispersions as a function of distance \label{tab:Dispersion_distance}}
\tablecolumns{8}
\tablenum{5}
\tablewidth{0pt}
\tablehead{
\colhead{Excluded} & \colhead{Radii} & \colhead{$\sigma_{\mu,\alpha^*}$} &
\colhead{$\sigma_{\mu,\delta}$} & \colhead{$\sigma_{\mu,r}$} & 
\colhead{$\sigma_{\mu,t}$} & \colhead{$\sigma_{\mu,1D}$}  & \colhead{N} \\
\colhead{ESC group} & \colhead{} & \colhead{} & \colhead{} & \colhead{} & \colhead{} & \colhead{} & \colhead{} \\
\colhead{} & \colhead{arcmin} & \colhead{mas yr$^{-1}$} & \colhead{mas yr$^{-1}$} &
\colhead{mas yr$^{-1}$} & \colhead{mas yr$^{-1}$} & \colhead{mas yr$^{-1}$} & \colhead{} \\
}
\startdata
 & $0.0 - 0.7$ & $1.20\pm0.10$ & $1.40\pm0.11$ & $1.26\pm0.10$ & $1.41\pm0.11$ & $1.30\pm0.07$ &91\\
 & $0.7 - 1.4$ & $0.93\pm0.06$ & $1.18\pm0.08$ & $1.07\pm0.07$ & $1.10\pm0.07$ & $1.06\pm0.05$ &153\\
None & $1.4 - 2.1$ & $0.80\pm0.06$ & $1.11\pm0.08$ & $0.96\pm0.07$ & $0.99\pm0.07$ & $0.97\pm0.05$ &135\\
 & $2.1 - 2.8$ & $0.80\pm0.05$ & $1.00\pm0.06$ & $0.99\pm0.06$ & $0.82\pm0.05$ & $0.91\pm0.05$ &158\\
 & $2.8 - 3.5$ & $0.81\pm0.07$ & $1.00\pm0.08$ & $0.83\pm0.07$ & $0.97\pm0.08$ & $0.91\pm0.05$ &93\\\hline
 & $0.0 - 0.7$ & $0.99\pm0.09$ & $1.40\pm0.11$ & $1.13\pm0.10$ & $1.30\pm0.11$ & $1.21\pm0.07$ &87\\
 & $0.7 - 1.4$ & $0.90\pm0.06$ & $1.18\pm0.08$ & $1.02\pm0.07$ & $1.04\pm0.07$ & $1.04\pm0.05$ &152\\
Group 1 & $1.4 - 2.1$ & $0.80\pm0.06$ & $1.11\pm0.07$ & $0.96\pm0.07$ & $0.99\pm0.07$ & $0.97\pm0.05$ &135\\
 & $2.1 - 2.8$ & $0.80\pm0.05$ & $1.00\pm0.06$ & $0.99\pm0.06$ & $0.82\pm0.05$ & $0.91\pm0.05$ &158\\
 & $2.8 - 3.5$ & $0.75\pm0.06$ & $1.00\pm0.08$ & $0.84\pm0.07$ & $0.92\pm0.08$ & $0.88\pm0.05$ &92\\\hline
 & $0.0 - 0.7$ & $0.98\pm0.09$ & $1.18\pm0.10$ & $1.04\pm0.10$ & $1.12\pm0.10$ & $1.08\pm0.07$ &79\\
 & $0.7 - 1.4$ & $0.90\pm0.06$ & $1.13\pm0.07$ & $0.99\pm0.07$ & $1.03\pm0.07$ & $1.02\pm0.05$ &149\\
Group 1 + 2 & $1.4 - 2.1$ & $0.80\pm0.06$ & $1.11\pm0.07$ & $0.96\pm0.07$ & $0.99\pm0.07$ & $0.97\pm0.05$ &135\\
 & $2.1 - 2.8$ & $0.80\pm0.05$ & $0.96\pm0.06$ & $0.96\pm0.06$ & $0.81\pm0.05$ & $0.88\pm0.04$ &157\\
 & $2.8 - 3.5$ & $0.75\pm0.06$ & $1.00\pm0.08$ & $0.84\pm0.07$ & $0.92\pm0.08$ & $0.88\pm0.05$ &92\\
\enddata
\end{deluxetable*}

\section{Discussion} \label{sec:Discussion}

 \begin{figure}
\plotone{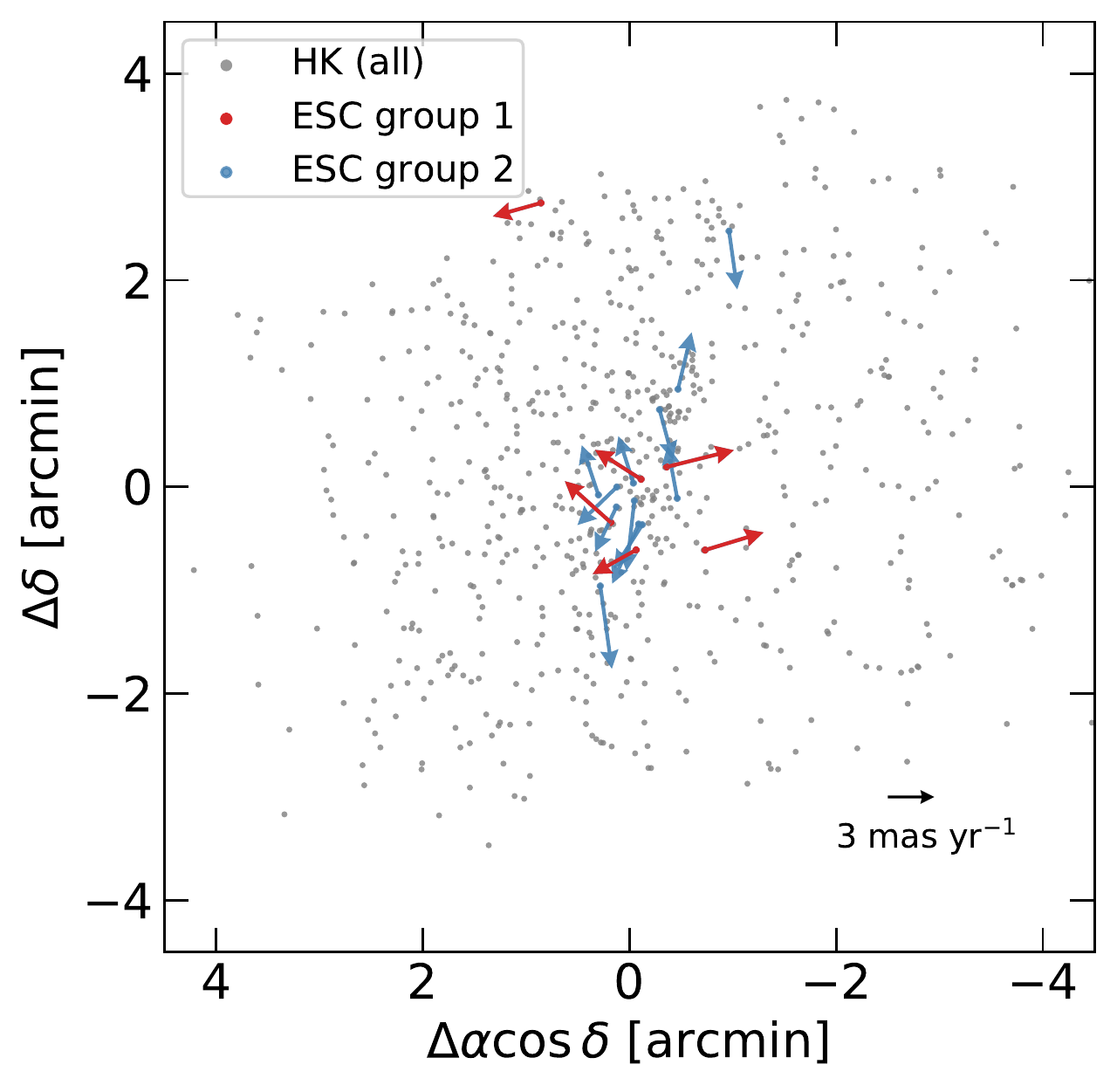}
\caption{The positions of stars with our \textit{HST}+Keck (HK) PM measurements are shown as grey points. Vectors represent the PMs of the escaping star candidates (ESC) in the right panel of Figure~\ref{fig:VPD}. 
\label{fig:outliers}}
\end{figure}

\begin{figure}[ht!]
\plotone{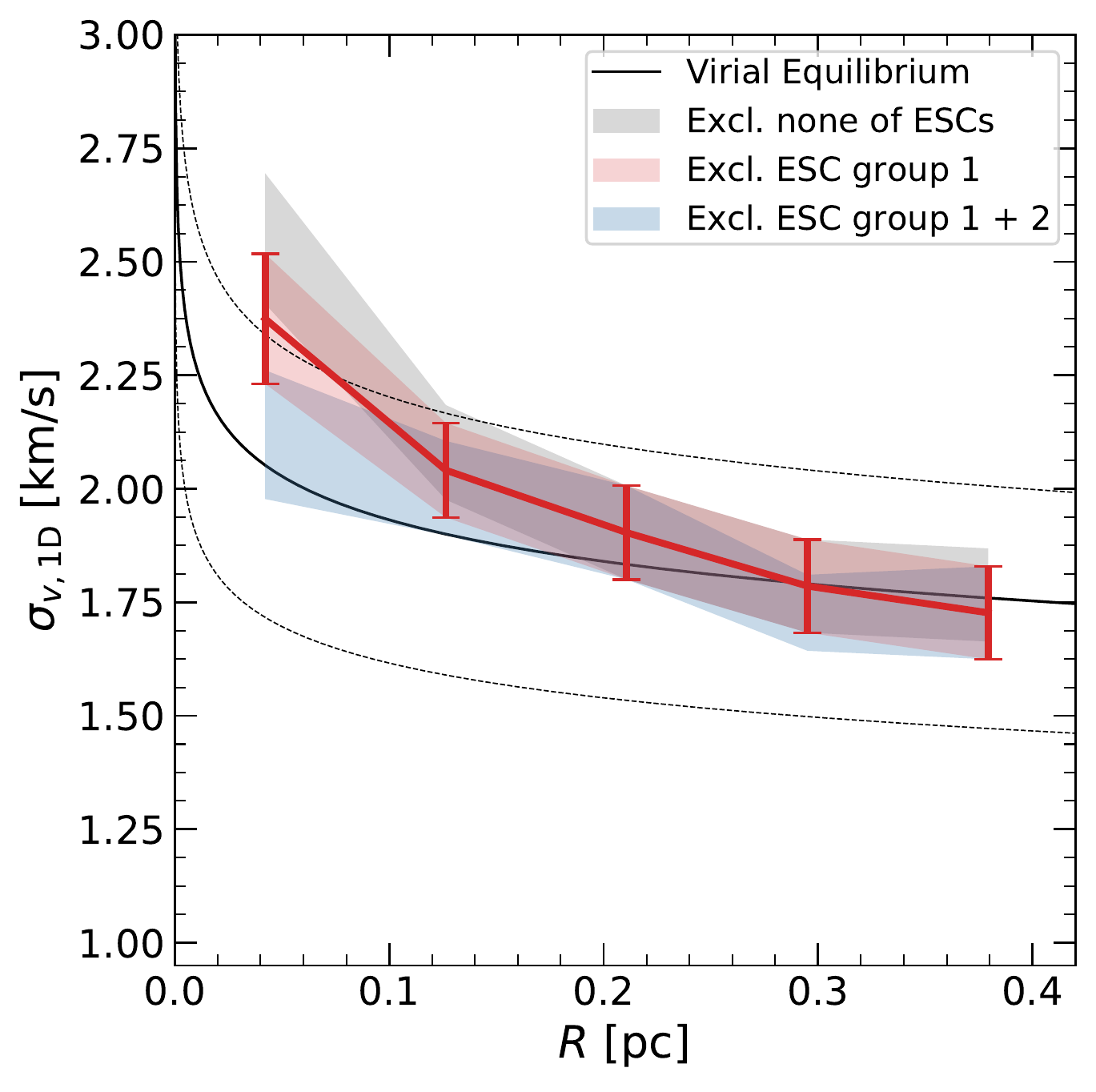}
\caption{Plot of the velocity dispersion $\sigma_{v,\mathrm{1D}}$ versus distance from the center of the ONC. The gray, red (with errorbars), and blue confidence bands mark velocity dispersions based on the PM dispersions presented in Table~\ref{tab:Dispersion_distance} and the estimate for the distance of the ONC, $414\pm7$ pc~\citep{Menten2007}. The black solid line illustrates the one-dimensional velocity dispersion for virial equilibrium predicted from the stellar and gas mass from~\cite{DaRio2014}, and the dashed lines mark the uncertainty assuming a 30\% mass uncertainty. Note that 0.3 pc corresponds to $\sim2\farcm5$ in radius. 
\label{fig:sigma_vs_radii}}
\end{figure}

\subsection{Normality and isotropy of PM distribution}
\label{sec:isotropy}

\cite{Kuhn2019} demonstrated that the ONC is one of only a few young open clusters whose stellar velocities are consistent with a multivariate normal distribution, or a thermodynamic Maxwell-Boltzmann distribution~\citep[see Figure 10 in][]{Kuhn2019}. For this analysis, however, the authors used \textit{Gaia} DR2 sources with \texttt{astrometric\_excess\_noise=0}, only a few of which fall in the central region of the ONC, as shown in Figure~\ref{fig:mosaic}. Their sample hence does not fully reflect the distribution of stellar velocities over the region covered in this work. In \S\ref{sec:high_vel_stars}, we identified deviation from normality at the tails of the distribution. To verify the multivariate normality more quantitatively, we employed the \texttt{R} package \texttt{MVN}~\citep{MVN} based on the method of \cite{HZ1990}.As in \S\ref{sec:high_vel_stars}, we tested the three cases: excluding (a) none of ESC groups, (b)  ESC group 1, and (c)  ESC group 1 + 2. The first case exhibits statistically significant deviation from normality ($p\sim0.01$) while the latter two cases show consistency with a multivariate normal distribution ($p>0.05$). The deviation from normality suggests that including the ESCs in ESC group 1 would overestimate the width of the PM distribution. 

Understanding the nature of the kinematic outliers is  important for assessing the applicability of our PM distribution model used in the previous section. We notice in Figure~\ref{fig:outliers} that these stars are mostly concentrated at the cluster core~\citep[$r_c\sim0\farcm1$; ][]{HH1998}, with PM vectors heading outward. This is also the case for the stars in ESC group 2. Their positions and motions imply that the majority are likely higher-velocity stars escaping the ONC as a consequence of more frequent dynamical interactions between stars at the cluster core~\citep{Johnstone1993,Baumgardt2002}. Another possibility is that some of these are unresolved binaries centrally concentrated due to mass segregation, although in this case, their anisotropic radial PMs would not be easily explained.
It is yet difficult to take a complete census of escaping stars due to measurement errors induced by the strong nebulosity and the lack of line-of-sight velocity measurements. 
With the relation $\langle v^{2} \rangle$ = $\sigma_{v}^{2}$ for the Maxwellian distribution and the PM dispersions calculated in \S\ref{sec:dispersion}, Equation (1) would give an estimate for the mean-square escape speed of $2.8\pm0.1$ mas yr$^{-1}$, $\sim10$\% lower than our previous estimate in \S\ref{sec:high_vel_stars}. 
This suggests that the previous estimate for the escape velocity needs to be treated as an upper limit, and that applying the lower limit could reveal additional candidates. Nonetheless, if the higher velocity stars are not escaping, then the deviation from normality seen in case (a) would be attributed to the rapid variation of velocity dispersions within the core of the cluster as shown in Figure~\ref{fig:sigma_vs_radii}. We note that, even when including the ESCs, the PMs of all five subsamples grouped by distance in Table~\ref{tab:Dispersion_distance} are consistent with a multivariate normal distribution ($p>0.05$) in both the Cartesian and radial-tangential coordinates. Our model is thus still valid in all the radial bins.

The ONC is well known to have a stellar velocity distribution that is elongated north-south~\citep[e.g.][]{JW1988,Kuhn2019} along the axis of the Orion A cloud filament. 
The PM distribution of our sample also appears elongated north-south with an axis ratio $\sigma_{\mu,\alpha^*}/\sigma_{\mu,\delta} = 0.74\pm0.03$. This kinematic anisotropy agrees with that seen in the stellar distribution at a radius of $\sim3\arcmin$, which is also elongated north-south with $b/a\sim0.7$~\citep[see Figure 3 in ][]{DaRio2014}, parallel to the Orion A molecular cloud~\citep{HH1998}. The PM dispersions may reflect the initial conditions of the protocluster cloud or the geometry of the present-day gravitational potential. On the other hand, the deviation from tangential-to-radial isotropy \citep[$\sigma_{\mu,t}/\sigma_{\mu,r}-1$;][]{Bellini2018} is $0.03\pm0.04$, which suggests that the cluster is consistent with being isotropic in the tangential-radial velocity space.

\subsection{Dynamical equilibrium}

It has been argued in some previous studies that the ONC is likely to be supervirial~\citep[e.g.,][]{JW1988, DaRio2014}. The dynamical mass of the ONC inferred from the previous kinematic analysis done by \cite{JW1988} is nearly twice the total stellar and gas mass. 
Alternatively, virial equilibrium requires a one-dimensional mean velocity dispersion $\sigma_{v}\simeq1.7\pm0.3$\,km s$^{-1}$ given the volume density of the stellar and gas contents in the ONC~\citep{DaRio2014}, which is only $\sim75\%$ of the velocity dispersion of $2.34\pm0.06$\,km s$^{-1}$ from \cite{JW1988}, leading to the conclusion that the ONC is likely to be slightly supervirial, with a virial ratio of $q\simeq0.9\pm0.3$. 

This past result is, however, partially attributable to the previous estimate of the distance of the ONC adopted for the derivation of velocity dispersion in \cite{JW1988}, $\sim470$ pc. 

The estimate of the distance has subsequently decreased in later studies to $\sim400$\,pc~\citep[e.g.,][]{Menten2007,Jeffries2007,Kounkel2017,Kounkel2018,Kuhn2019,Grossschedl2018}. At a distance of $d=414\pm7$ pc from~\cite{Menten2007}, \cite{JW1988} would have obtained a smaller value for their velocity dispersion $\sigma_{v}\simeq2.1\pm0.1$ km s$^{-1}$, which would give a virial ratio as low as $q\simeq0.7\pm0.3$.

Figure~\ref{fig:sigma_vs_radii} compares our measured velocity dispersions in Table~\ref{tab:Dispersion_distance} to the predicted velocity dispersions required for virial equilibrium based on the total mass profile from \cite{DaRio2014}. We note that we adopted $414\pm7$ pc from~\cite{Menten2007} for the distance of the ONC as \cite{DaRio2014} did, for consistency purposes. Overall, our velocity dispersion profiles tend to decrease with radius following the prediction based on the observed mass profile, a power-law profile slightly steeper than a singular isothermal sphere~\citep{DaRio2014}. 
When we include the escaping star candidates, our measured velocity dispersion appears to be more than $1\sigma$ larger than predicted at the very central region. However, when neglecting the escaping stars in ESC group 1 or 1 + 2, our measurements are in good agreement with the predicted values within $1\sigma$ uncertainty, suggesting that the bulk of the cluster is virialized. Even if the escaping star candidates are included, the measured velocity dispersions are still well below the boundedness limit ($\sigma_{bound}=\sqrt{2}\sigma_{vir}$). There is no indication of global expansion in the ONC from our analysis; the mean PM along the radial axis is small and consistent with zero within the $\sigma$ uncertainty, as shown in \S\ref{sec:dispersion}. 

\cite{Kuhn2019} has reported evidence of mild expansion in the ONC based on PM measurements of \textit{Gaia} DR2 sources, finding the median outward velocity $\widetilde{v_{out}}=0.42\pm0.20$ km$^{-1}$ based on the uncertainty-weighted median with bootstrap resampling. In Figure 5 of their paper, the weighted kernel-density estimate (KDE) plot of $v_{out}$ exhibits two peaks, one at $\sim-0.4$ km s$^{-1}$ and the other at $\sim1.0$ km s$^{-1}$, which implies a significant concentration of data points or weights at two different places. To explore this, we divided the \textit{Gaia} DR2 sample of \cite{Kuhn2019}, including 378 stars, into two subgroups in terms of weights: (a) 42 stars with small errors (i.e. large weights) $\epsilon_{v,out}<0.15$ km s$^{-1}$ and (b) 336 stars with larger errors $\epsilon_{v,out}>0.15$ km s$^{-1}$. The sum of the weights of the group (a) reaches $\sim64$\% of that of the group (b). For the groups (a) and (b), the median velocity is found to be $\widetilde{v_{out}}=0.94\pm0.41$ km $s^{-1}$ and $0.09\pm0.19$ km $s^{-1}$, respectively. We note that we converted PMs into $v_{out}$ and calculated the median velocity in the same manner as described in \S3 and \S4 of \cite{Kuhn2019}, for consistency purposes. This result indicates that the $\sim10$\% of the whole sample (i.e. group (a)) produces the second peak in the KDE plot and biases the median velocity, ultimately leading to the conclusion that the ONC shows evidence for mild expansion. In fact, the stars in group (a) mostly fall into the magnitude range between \texttt{phot\_g\_mean\_mag} $\sim13$ and 15, which corresponds to the transition point in the \textit{Gaia} error terms from the detector and calibration-dominated regime to photon noise-limited regimes. In this interval, the astrometric uncertainties of \textit{Gaia} DR2 are known to be the most underestimated~\citep[see \S4.6.4 and Figure 24 in][]{Arenou2018}. We refer to the IAU GA presentation slides by Lindegren\footnote{https://www.cosmos.esa.int/web/gaia/dr2-known-issues} for more details. Between the group (a) and our catalog, we found one star matched, whose PM errors in the \textit{Gaia} DR2 need to be increased by a factor of $\sim3$ or greater for its outward velocities to be consistent within 1$\sigma$ uncertainty (see also \S\ref{sec:ConsistencyWithDR2}). We therefore conclude that the net outward velocity claimed by \cite{Kuhn2019} is likely a result of underestimated uncertainties in the \textit{Gaia} DR2 data.

Our result adds to the growing evidence that the central region of the ONC is dynamically evolved. Studies of spatial morphology have revealed that the ONC has overall very little stellar substructure~\citep{HH1998,DaRio2014}. The core of the cluster exhibits a rounder and smoother stellar distribution than the outskirts, indicating that the core has likely experienced more dynamical timescales to lose initial substructures. The line-of-sight velocities are also smoothly distributed, as expected from an old dynamical age~\citep{DaRio2017}. Correcting for the local variation of the mean velocities, \cite{DaRio2017} found that the dispersion of the line-of-sight velocities is measured as low as $\sim1.7$ km s$^{-1}$, which agrees with a virial state. \cite{Kuhn2019} has also reached a similar conclusion using the PMs of 48 \textit{Gaia} DR2 sources. 

The dynamical age also has an implication for the origin of mass segregation. 
The ONC exhibits clear evidence of mass segregation, with the most massive stars preferentially located in the central region of the cluster~\citep{HH1998}. The young age of the stellar population, as young as $\sim2.5$ Myr on average, is often cited as evidence for the primordial origin of the mass segregation. The evidence of virial equilibrium, however, implies that the central region of the ONC is dynamically old enough to have undergone several crossing times or more~\citep{Tan2006}. With a large age spread of $\sim1.3$ Myr, the stellar population already appears several times older than the current crossing time based on the observed cluster mass~\citep{DaRio2014}. In addition, the dynamical timescale was likely much smaller at the early phase of the cluster due to higher stellar densities~\citep{Bastian2008,Allison2009}. The mass segregation in the ONC thus need not be fully primordial.

Ultimately, our results favor the theoretical hypothesis that star cluster formation is a dynamically ``slow'' process; stars form slowly in supersonically turbulent gas over several crossing times, reaching a quasi-equilibrium~\citep{Tan2006,KT2007,Krumholz2012}. In the ``fast'' scenarios, a star cluster forms via a rapid, global collapse of a gas clump within $\sim1$ crossing time~\citep[e.g.,][]{Elmegreen2007,Krumholz2011}, in which case the cluster would lack features expected from dynamical evolution. The ONC  continues to form stars with low efficiency and largely in virial equilibrium, and thus provides observational evidence for the slow formation scenario and therefore does not support the competitive accretion model.

\begin{figure}[ht!]
\plotone{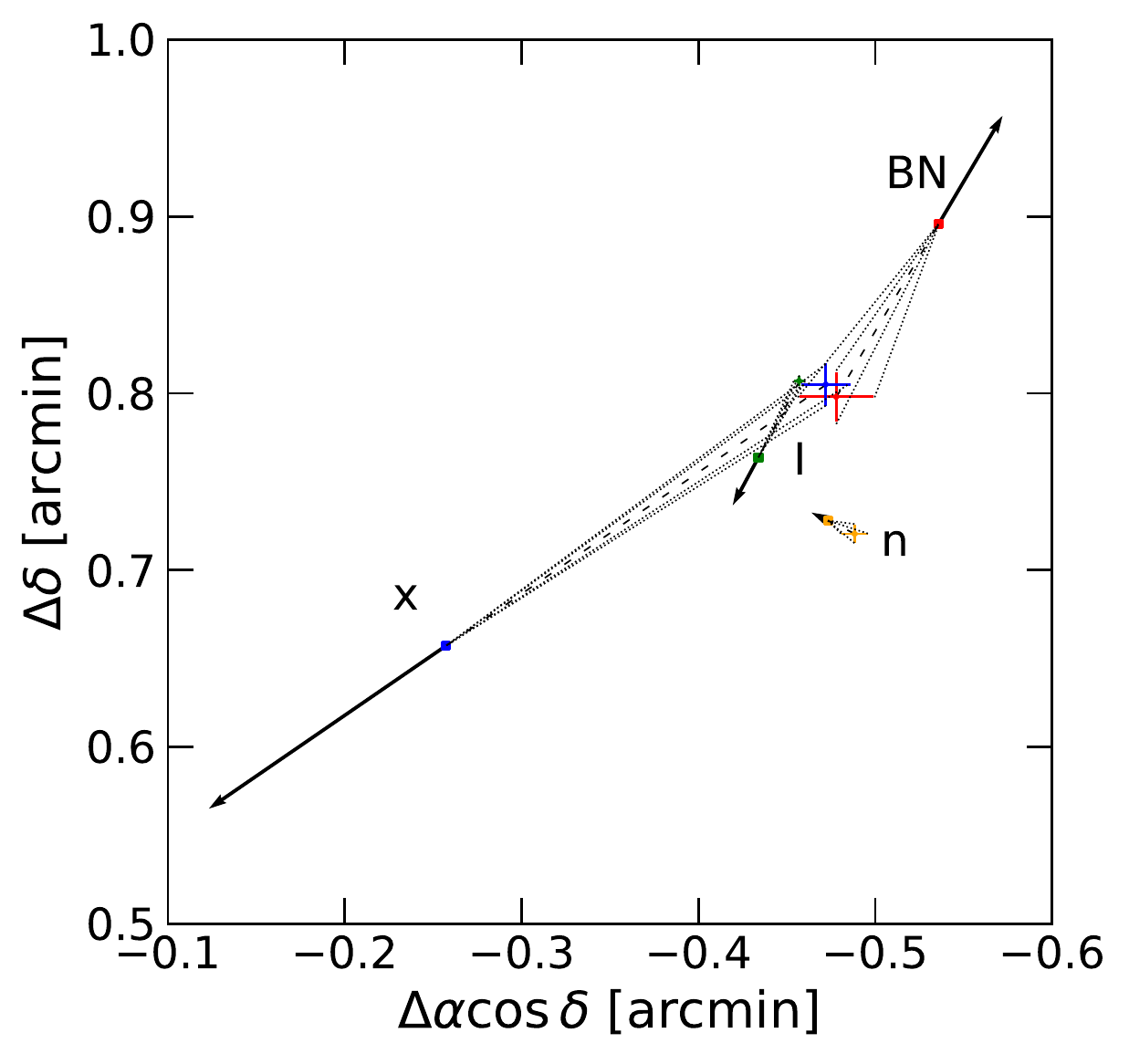}
\caption{Positions of BN, source \textit{I}, source \textit{n}, and source \textit{I} for the epoch of 2015 (squares without errorbars). Arrows indicate the direction and PM displacement for 300 years in the rest frame of ONC. The dashed and dotted lines illustrate the projected paths back to 1535 and uncertainties, respectively. The errorbars represent the estimated positions for the epoch of 1535. The astrometry and PM of source \textit{I} were adopted from~\cite{Rodriguez2017}. 
\label{fig:HighVelocityStars}}
\end{figure}

\subsection{Origin of high-velocity stars}

Our PM catalog includes previously known fast-moving sources around the BN/KL complex, namely BN, source \textit{I}, and source \textit{n}, although in our analysis source \textit{n} exhibits a rather small PM in the rest frame of the ONC. These optically invisible objects are all detected in epoch 2010 and 2014 NIRC2 He I b ($\sim$ K-band) images\footnote{The astrometric measurements of the highly embedded object BN on the HST WFC3/IR F130N and F139M ($\sim$ \textit{J} and \textit{H}-bands) images were excluded from this analysis as it appears highly elongated and asymmetric on those images.}. Before the recent discovery of source \textit{I}, two possible scenarios were proposed to explain their origin:

(a) Ejected from the Trapezium $\theta^{1}$ Ori C $\sim4000$ years ago~\citep{Plambeck1995,Tan2004}, BN passed near  source \textit{I} triggering an explosive outflow or (b) BN, source \textit{I}, and at least one other star once comprised a multiple system, and the latter two objects merged into a tight binary system, which resulted in an explosive outflow $\sim500$ years ago~\citep{Rodriguez2017}. 

Numerous pieces of evidence argue against the first scenario, e.g., the large separation between $\theta^{1}$ Ori C and BN at the time of ejection ($\gtrsim10''$) and their inconsistent momenta and ages~\citep[see][and references therein]{Goddi2011}. In the meantime, the source \textit{I} was recently found as a promising candidate of the formerly missing puzzle for the second scenario. \cite{Luhman2017} demonstrated that the estimated location of source \textit{I} $\sim500$ years ago agrees with the position for BN and source \textit{I} at that time, determined by radio PMs measured by~\cite{Gomez2008,Goddi2011,Rodriguez2017}, which suggests that the three sources were likely ejected in the same event. 

We have independently calculated when BN and source \textit{I} experienced their closest approach ($t_{min}$) using our PM measurements and the method described in Appendix B. We obtained $t_{min}=1535\pm29$ yr, which is consistent within $2\sigma$ with that for BN and source \textit{I} based on their radio PMs, $t_{min}=1475\pm6$ yr~\citep{Rodriguez2017}. Figure~\ref{fig:HighVelocityStars} shows the PM vectors and the positions of BN, source \textit{I}, \textit{n}, and \textit{I} and $1\sigma$ range of allowed paths back to 1535, where only the data for source \textit{I} have been adopted from~\cite{Rodriguez2017}. Our observation supports the possibility that BN, source \textit{I}, and \textit{I} were ejected from around the same location $\sim500$ years ago. Elements of this hypothesis were once challenged by theoretical simulations; \cite{Farias2018} demonstrated using a large suite of \textit{N-body} simulations that the mass for the source \textit{I} required by their simulations is as large as at least 14$M_{\odot}$, almost as large as than the previously estimated $\sim7\,M_{\odot}$ based on the kinematics of the circumstellar material~\citep{Matthews2010,Hirota2014,PW2016}. Recent ALMA observations with higher resolution and sensitivity than the previous ones, however, estimate the mass of source \textit{I} as $15\pm2M_{\odot}$~\citep{Ginsburg2018}, by which the dynamical decay scenario still remains viable.

\section{Conclusions} \label{sec:Conclusions}

Using \textit{HST} ACS/WFC3IR data that span $\sim20$ years and Keck II NIRC2 data obtained in 2010 and 2014, we obtained relative PMs of 701 stars within $\sim3\farcm0$ of the ONC. With the analysis of these PMs, we reach the following conclusions.

\begin{enumerate}
\item Excluding the kinematic outliers, the PMs of our sample are consistent with a multivariate normal distribution. With the refined sample, the calculated velocity dispersions are $(\sigma_{\mu,\alpha^*},\,\sigma_{\delta})=(0.83\pm0.02,\,1.12\pm0.03)$ mas yr$^{-1}$ and $(\sigma_{\mu,r},\,\sigma_{\mu,t})=(0.97\pm0.03,\,1.00\pm0.03)$ mas yr$^{-1}$. These values agree with those in previous surveys~\citep[e.g.,][]{JW1988}; but have a factor of $\sim2$ improved precision.

\item The PM distribution appears elongated north-south with an axis ratio of $\sigma_{\mu,\alpha^*}/\sigma_{\mu,\delta}=0.74\pm0.03$, resembling the stellar distribution which is also elongated north-south with $b/a\approx0.7$ in the central region. On the other hand, the radial and tangential PMs are consistent with tangential-to-radial isotropy, as indicated by the low deviation from isotropy ($\sigma_{\mu,t}/\sigma_{\mu,r}-1)=0.03\pm0.04$.

\item Compared to the prediction from the total density profile, our velocity dispersion profile is in good agreement with a virialized state. This suggests that the star-forming region is dynamically evolved. 

\item Our analysis recovered the fast-moving IR sources in the BN/KL region, including BN, \textit{I}, and \textit{n}. The PMs of BN and source \textit{I} are consistent with previous measurements in the literature, whereas source \textit{n} exhibits a relatively small PM, as previously seen at IR and millimeter wavelengths. The estimated locations of BN and source \textit{I} when the closest separation took place agree with the initial position of the radio source \textit{I}, implying the dynamical decay of a multiple system involving these three sources.

\item The majority of ESCs are concentrated around the core of the ONC, where their PM vectors mostly point outward. 

\item Based on comparisons with current star formation theories, our result suggests that the ONC is forming stars with a low star formation efficiency per dynamical timescale.

\end{enumerate}

Our analysis shows that high spatial resolution, near-IR coverage of the ONC is essential; \textit{HST} WFC3/IR + Keck NIRC2 observations revealed a factor of $\sim3$ more stars than the optical \textit{Gaia} DR2 sources in the BN/KL region. In order to obtain a complete PM catalog of the most embedded and the lowest mass objects, additional observations over a sufficiently long time baseline with near-IR telescopes/instruments such as \textit{HST} WFC3/IR and Keck NIRC2 and the next generation near/mid-IR telescopes such as \textit{JWST} and \textit{WFIRST} will be required. Alongside the PM analysis, ongoing spectroscopic surveys for stellar line-of-sight velocities around the ONC will enable determining 3-dimensional stellar velocity distribution (e.g., Theissen et al. in preparation).

\acknowledgments\noindent \textbf{Acknowledgments.}  
The authors thank the anonymous referee for interesting suggestions and comments. We thank Andrea Bellini, Joshua Simon, and Anthony Brown for helpful discussions on \textit{Gaia} data, Gaspard Duchene on visual binaries in the ONC, and Michael Kuhn for providing the list of \textit{Gaia} DR2 source IDs. We also thank Steven Stahler and Lynne Hillenbrand for informative discussions and comments. J.~R.~L. and D.~K. acknowledge support from NSF AST-1764218, HST-AR-13258, and HST-GO-13826. J.~R.~L., Q.~K., D.~K., and C.~A.~T. are supported by the NSF grant AST-1714816. This work is based on observations made with the NASA/ESA Hubble Space Telescope, obtained at the Space Telescope Science Institute, which is operated by the Association of Universities for Research in Astronomy, Inc., under NASA contract NAS 5-26555. 
This work has made use of data from the European Space Agency (ESA) mission {\it Gaia} (\url{https://www.cosmos.esa.int/gaia}), processed by the {\it Gaia} Data Processing and Analysis Consortium (DPAC, \url{https://www.cosmos.esa.int/web/gaia/dpac/consortium}). Funding for the DPAC has been provided by national institutions, in particular the institutions participating in the {\it Gaia} Multilateral Agreement.
Some of the data presented herein were obtained at the W. M. Keck Observatory, which is operated as a scientific partnership among the California Institute of Technology, the University of California and the National Aeronautics and Space Administration. The Observatory was made possible by the generous financial support of the W. M. Keck Foundation.
The authors wish to recognize and acknowledge the very significant cultural role and reverence that the summit of Maunakea has always had within the indigenous Hawaiian community.  We are most fortunate to have the opportunity to conduct observations from this mountain.

We acknowledge use of the Astropy \citep{astropy}, Matplotlib \citep{matplotlib}, and R \citep{RCoreTeam2018} software packages. 

\facilities{HST(ACS, WFPC2, WFC3), Keck:II (NIRC2), Gaia}
\software{Astropy, align, emcee, hst1pass, img2xym\_WFC, img2xymrduv, Matplotlib, MVN, R, StarFinder}

\appendix

\section{Systematic Errors in \textit{Gaia} DR2 PMs}

\begin{figure}[ht!]
\plotone{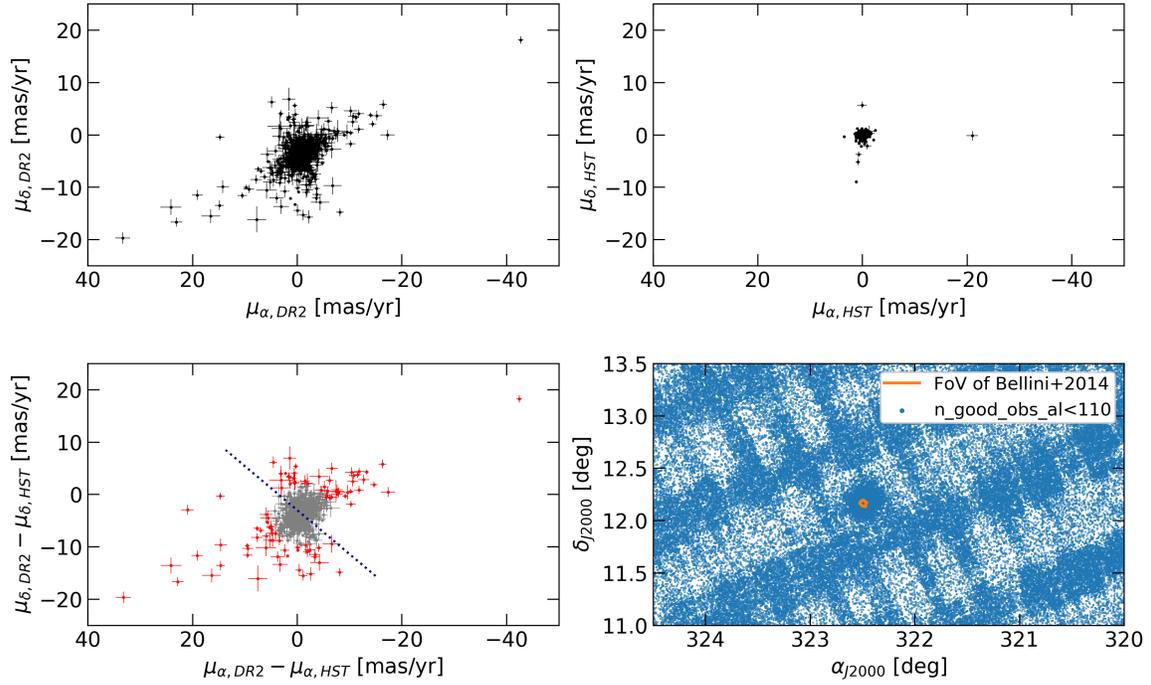}
\caption{Evidence of systematics in \textit{Gaia} DR2 PMs. Top left panel: the distribution of \textit{Gaia} DR2 PMs for stars in common between \textit{Gaia} DR2 and the \textit{HST} NGC\,7078 PM catalog of \cite{Bellini2014}. Top right panel: the same as the top left panel but for HST PMs.  Bottom left panel: the same as the top panels but for differences in the \textit{Gaia} DR2 and \textit{HST} PMs. Obvious outliers are highlighted in red for Figure~\ref{fig:ngc7078b}. The dotted line illustrates the orientation of the Galactic plane. Bottom right panel: the positions of \textit{Gaia} DR2 sources with \texttt{astrometric\_n\_good\_obs\_al} below 110, which imprint the scanning law of \textit{Gaia}. The orange polygon marks the field of view of the \textit{HST} catalog.
\label{fig:ngc7078a}}
\end{figure}

In order to use the \textit{Gaia} DR2 as a control sample (see \S\ref{sec:ConsistencyWithDR2}), it is important to understand underlying systematic errors as well as random errors depending on the quality of individual measurements. Here, we demonstrate that the \textit{Gaia} DR2 PMs may include systematic errors induced by the scanning law of the survey, by comparing to \textit{HST} PMs in the globular cluster NGC\,7078. Note that NGC\,7078 provides a cleaner data set than the nebulous ONC. The top left and right panels of Figure~\ref{fig:ngc7078a} show the PM vector point diagrams of stars in common between \textit{Gaia} DR2 and the HST PM catalog of \cite{Bellini2014}. The distribution of \textit{Gaia} DR2 PMs exhibits unexpected linear structures nearly perpendicular to one another. These structures still remain in the bottom left panel, where the \textit{HST} PMs are subtracted from the \textit{Gaia} DR2 PMs. The linear trends closely resemble the scanning footprints around the cluster, which can be traced by the positions of \textit{Gaia} DR2 sources filtered based on the number of good observations, \texttt{astrometric\_n\_good\_obs\_al}, as shown in the bottom right panel. This suggests that the systematic errors reflect the scanning pattern and survey incompleteness. 

\begin{figure}[ht!]
\plotone{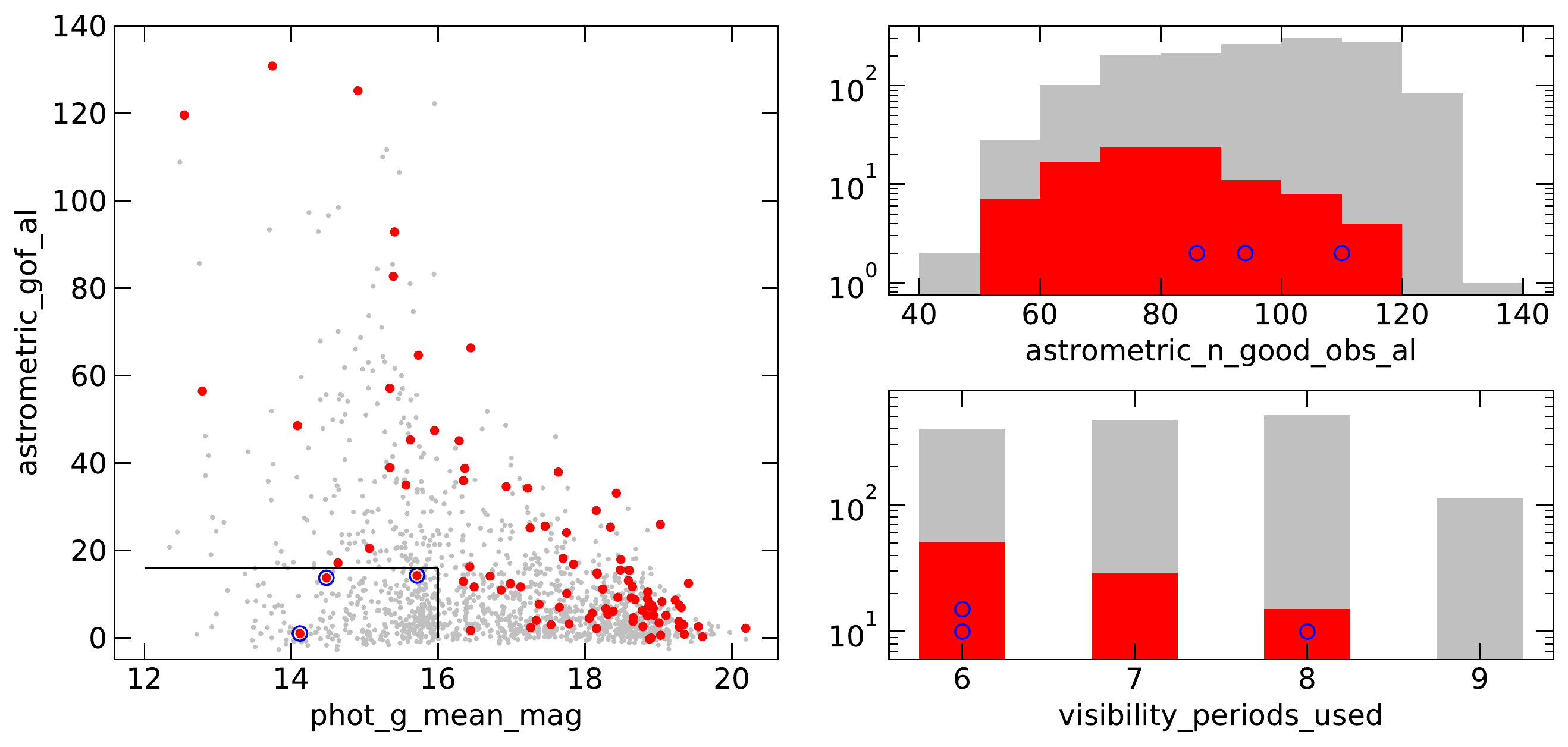}
\caption{Characteristics of \textit{Gaia} DR2 sources with systematic errors. Filled circles in red and gray correspond to the selected outliers and the rest in the bottom left panel of Figure~\ref{fig:ngc7078a}, respectively. Left panel: goodness-of-fit statistic as a function of magnitude. Black solid lines outline the quality cut thresholds applied for our control sample in \S\ref{sec:ConsistencyWithDR2}. Outliers that passed the cuts are marked with empty circles in blue. Right panels: histograms of \texttt{astrometric\_n\_good\_obs\_al} (top) and \texttt{visibility\_periods\_used} (bottom).
\label{fig:ngc7078b}}
\end{figure}

We find that fainter stars with a smaller number of observations and visibility periods tend to be more strongly effected by systematics. Figure~\ref{fig:ngc7078b} compares obvious outliers along the structures, highlighted in red in the left bottom panel of Figure~\ref{fig:ngc7078a}, to the rest in different parameter spaces. The outliers are mostly fainter than \texttt{phot\_g\_mean\_mag}$\sim16$ and have \texttt{astrometric\_n\_good\_obs\_al} values below 120 and \texttt{visibility\_periods\_used} values below 8. The black solid lines in the left panel mark the quality cut thresholds applied for our control sample in \S\ref{sec:ConsistencyWithDR2} based on magnitudes and goodness-of-fit (gof) statistic of individual sources. These criteria alone cannot completely rule out the possibility of systematic errors; there are still a few contaminants, circled in blue, inside the boundaries. The right panels show that increasing thresholds for \texttt{astrometric\_n\_good\_obs\_al} and \texttt{visibility\_periods\_used} can reduce the number of stars subjected to the systematic errors in a control sample. Filtering based on these two parameters may come at a cost, especially where the detection efficiency of \textit{Gaia} is typically low like at the central region of globular clusters due to high crowding~\citep{Arenou2018}. 

\section{The minimum separation of BN and source x in the past}

We have determined the closest approach distance between BN and source \textit{I} based on our PM measurements in a similar approach to the method of~\cite{Gomez2008}. Assuming their PMs are linear, the relative positions of BN with respect to source \textit{I} as a function of time $t$ are given by 
$$ x(t) = x_{2014.9} + \mu_x(t-2014.9),$$
$$ y(t) = y_{2014.9} + \mu_y(t-2014.9),$$
where ($x_{2014.9}$, $y_{2014.9}$) and ($\mu_x, \mu_y$) are relative positions for epoch 2014.9 and relative motions in the right-handed Cartesian coordinates (x=$\Delta\alpha\cos\delta$, y=$\Delta\delta$). The separation of the two sources as a function of time is then given by

$$s(t) = \sqrt[]{x^2(t)+y^2(t)} $$

The minimum separation $s_{min}$ and the corresponding epoch $t_{min}$ are given by differentiating the separation and equaling to 0:
$$s_{min}=\frac{|x_{2014.9}\mu_x-y_{2014.9}\mu_y|}{\sqrt[]{\mu^2_x+\mu^2_y}}$$
$$t_{min}=-\frac{x_{2014.9}\mu_x+y_{2014.9}\mu_y}{\mu^2_x+\mu^2_y}$$
Given a relative position of ($x_{2014.9}$, $y_{2014.9}$)= ($-16706.2$, 14309.9) mas and PMs of ($\mu_x, \mu_y$) = ($-34.0\pm3.1$, $30.6\pm2.4$) mas yr$^{-1}$  based on our measurements for BN and source \textit{I}, we obtain

$$ s_{min} = 2\farcs82 \pm 1\farcs32,$$
$$ t_{min} = \,1535\pm29 \;\textrm{yr}.$$



\end{document}